\documentclass[amsmath,amssymb,floatfix,showpacs,twocolumn]{revtex4}
\input epsf
\newcommand {\eq}{\begin{equation}}
\newcommand {\qe}{\end{equation}}

\newcommand {\bfq} {{\bf q}}
\newcommand {\bfr} {{\bf r}}

\newcommand {\bfb} {{\bf b}}

\newcommand {\bfR} {{\bf R}}
\newcommand {\bfQ} {{\bf Q}}

\newcommand {\bfu} {{\bf u}}

\newcommand {\h}{\frac{1}{2}}

\newcommand {\bfs}{{\bf s}}

\newcommand {\ar}{a_R}
\newcommand {\ai}{a_I}

\newcommand {\nucp}{Nucl. Phys.}

\usepackage{epsfig}
\begin{document}
\vspace*{-1.0in}


\title {Monte Carlo Eikonal Scattering}

\author{W. R. Gibbs}

\affiliation{\large New Mexico State University, Las Cruces,
NM 88003}

\author{Jean-Pierre Dedonder}

\affiliation{\large Laboratoire de Physique Nucl\'{e}aire et Hautes 
\'{E}nergies\\
Universit\'es Pierre et Marie Curie et Paris-Diderot, IN2P3 et CNRS\\
4, place Jussieu, 75252 Paris cedex 05, France}

\date{\today}

\begin{abstract}

\begin{description}\item[Background]The eikonal 
approximation is commonly used to calculate heavy-ion 
elastic scattering. However, the full evaluation has only 
been done (without the use of Monte Carlo techniques or additional 
approximations) for $\alpha-\alpha$ scattering.

\item[Purpose]Develop,  improve and test the Monte Carlo 
eikonal method for elastic scattering over a wide range of 
nuclei, energies and angles.

\item[Method] Monte Carlo evaluation is used to calculate 
heavy-ion elastic scattering including the center-of-mass 
correction and the Coulomb interaction.

\item[Results] Angular distributions are presented for a
number of nuclear pairs over a wide energy range using 
nucleon-nucleon scattering parameters taken from phase-shift 
analyses and densities from independent sources. A technique 
for the  efficient expansion of the Glauber amplitude in 
partial waves is developed.

\item[Conclusion] The center-of-mass and Coulomb corrections
are essential. Angular distributions can be predicted only up to
certain angles which vary with the nuclear pairs but all 
correspond to a momentum transfer near 1 fm$^{-1}$. 

\end{description}

\end{abstract}
\pacs{25.70.Bc,25.55.Ci}
\maketitle

\section{Introduction}

Monte Carlo (MC) methods are very valuable for performing 
multi-dimensional integrals, especially in quantum 
applications \cite{lhe}. Recently, calculations of nuclear 
structure using MC have been performed, both for 
light nuclei \cite{carlson,pudliner} and with the nuclear shell 
model \cite{mcsm}.  In the present article we study the use of this 
technique for the evaluation of multi-dimensional integrals in the 
calculation of elastic scattering of heavy ions in the 
eikonal approximation.

Monte Carlo techniques have been used for {\it classical} 
simulations of nuclear reactions for many years (see e.g. 
Ref. \cite{cexd} and references within).  In that case 
(highly inelastic reactions), the assumption that phase 
information is unimportant appears to be valid, at least 
under certain conditions. This technique can reveal 
properties of nuclear reactions for inelastic reactions where 
many channels are open. Much less use has been made of the MC 
method for elastic (or nearly elastic) reactions. The reason 
for this might be that integrals performed with Monte Carlo 
are deemed not to be suitable to produce the highly 
oscillating amplitudes because of the growth in errors that 
typically occurs in this case. As we shall see shortly the 
evaluation of the Glauber elastic scattering amplitude does 
not suffer from this problem.
We now review some of the literature on this problem.

\subsection{Review}

\subsubsection{The use of the Glauber approximation for nucleus-nucleus 
scattering}

Before discussing the use of Monte Carlo for eikonal 
calculations, we briefly mention the applications of the Glauber 
scattering method to heavy ion scattering more generally.

Franco and Yin \cite{fy} and Yin {\it et al.} \cite{yin} were 
the first to evaluate the 
full sum for the $\alpha-\alpha$ scattering amplitude which 
appears 
to be the largest case for which the full scattering 
amplitude can be evaluated without further approximation and 
without the use of Monte Carlo. Their method relies on the 
use of a Gaussian approximation for the nuclear density.
Abu-Ibrahim et al. \cite{abu5} evaluated the full Glauber 
multiple-scattering amplitude for proton-$^6$He scattering.

Al-Khalili, Thompson and Tostevin \cite{att} studied the 
$^{11}$Li halo structure using eikonal methods, Al-Khalili 
and Tostevin \cite{at} studied the relation of reaction cross 
section to radii in the Glauber model, They also 
studied proton helium halo nuclei cite{alkhalili1} 
with Glauber scattering and in Ref. \cite{alkhalili}, with
Brooke, they treated $^{11}$B+$^{12}$C scattering with non-eikonal 
corrections, modifying the trajectory at lower energies.
El-Gogary \cite{elg3} treated cluster nuclei to evaluate the 
Glauber formula. 
El-Gogary et al. \cite{elg2} used an approximate center-of-mass 
correction with a double Gaussian form for the density.
Franco and Varma \cite{fv} treated the center-of-mass effects 
to several orders and compared (primarily) with total cross 
sections. El-Gogary et al. \cite{elg1} did calculations with an 
approximate center-of-mass correction and noted that a simplified 
treatment of center-of-mass effects is problematic.
Horiuchi et al. \cite{horiuchi} did a systematic study of reaction
cross sections using an improved expansion of the Glauber formula.
Zhong \cite{zhong} calculated nucleus-nucleus scattering in an 
$\alpha$-cluster model with double Gaussian forms for the 
density.
Charagi and Gupta \cite{charagi} calculated $^{16}$O scattering 
from several nuclei using the Optical Limit Approximation (OLA).
Abu-Ibrahim and Suzuki \cite{abu4} used the OLA and a phenomenological 
profile function to treat a large number of nuclei. Abu-Ibraham et 
al. \cite{abu2} calculated halo nuclei scattering in the 
Glauber model. 
Abu-Ibrahim and Suzuki \cite{abu} calculated nucleus-nucleus 
scattering at intermediate energies using corrections to the 
OLA. Abu-Ibraham and Suzuki \cite{abu3} studied the profile 
function in heavy ion scattering.
Franco and Nutt \cite{fn} treated short-range correlations in 
heavy ion scattering.
Lenzi et al. \cite{lenzi1,lenzi2} did a systematic treatment 
of heavy ion scattering in a OLA treatment of eikonal 
scattering.

\subsubsection{Monte Carlo}

The Monte Carlo method has been used to evaluate the eikonal 
amplitude on several occasions.

A. M. Zadorozhnyj, V. V. Uzhinsky and S. Y. Shmakov \cite{zador} 
did perhaps the first application of eikonal  Monte Carlo but no 
details are given.
Merino, Novikov and Shabelski \cite{merino} very recently 
compared methods for radius extraction. One of the methods is 
exact with Monte Carlo using the Metropolis algorithm \cite{metropolis}.
Alkhazov and Lobodenka \cite{alkh} calculated reaction cross 
sections for halo nuclei using the Monte Carlo method.
Shmakov, Uzhinskii and Zadorozhny \cite{shmakov} used Monte Carlo 
techniques for the generation of inelastic diagrams.
Krpic and Shabelski \cite{krpic} used the Metropolis 
algorithm to calculate elastic and inelastic scattering from 
several nuclei using a diagrammatic expansion. They neglected 
the center-of-mass correction.
Abu-Ibrahim and Suzuki \cite{abu6} studied low-energy 
$^6$He-$^{12}$C scattering using the optical eikonal 
potential evaluated with Monte Carlo.

Recently Varga et al. \cite{varga} have combined the Monte 
Carlo Green's Function method of solution of few-body 
problems with Glauber theory to calculate $\alpha-\alpha$ and 
halo-nuclei scattering with no approximation beyond those 
inherent in the basic eikonal theory. No doubt this method 
should be used in any case that an exact calculation of the 
nuclear structure is available but a much more common case is 
the scattering of heavier nuclei where such a solution is 
still for the future. In this case less ambitious 
approximations to the nuclear densities need to be used. We 
later will show calculations comparing the two methods and it 
would seem that the details (short-range correlations) make 
little difference.

\subsubsection{Center of Mass}

Of the above methods only that of Varga et al. \cite{varga} 
includes the center-of-mass correction exactly. We will see that this 
correction is very important, even for heavier nuclei. 

Chauhan and Khan \cite{chauhankhan} treated 
$^{12}$C--$^{12}$C elastic scattering and found that 
center-of-mass effects play an important role. They use an 
expansion of the profile function.
Liu et al. \cite{liu} calculated $\alpha$ scattering from 
several nuclei at 1.37 GeV. They used an overall factor for 
the center-of-mass correction and varied the phase of the 
nucleon-nucleon interaction.

Franco and Tekou \cite{ft} calculated the optical model version
up to 5th order. They included corrections for the center of mass.
Shukla \cite{shukla} included the center-of-mass and  
Coulomb effects to investigate reaction cross sections.

\subsubsection{Coulomb}

The basic Glauber method does not include the Coulomb 
interaction and a correction (often very important) must be 
made for it after the calculation.

Kondratyuk and Kopeliovich \cite{kk} argued that it would be 
a good approximation to simply add the strong and Coulomb 
phases. Glauber and Matthiae \cite{matthiae} and Czyz, 
Lesniak and Wolek \cite{czyz} used mainly this approximation. 
The corrections to this approximation were developed using 
various techniques but most often based on an optical 
potential. See Ref. \cite{ga} for a recent treatment and 
references to the previous work for the application to 
pion-nucleon scattering.
F\"aldt and Pilkuhn \cite{faldt} used a semi-classical method 
with the Glauber model to correct for the Coulomb 
modification of the trajectories. This technique has the 
advantage that an optical model fit does not have to be made 
to the data to find the correction factors. While they developed 
the method for pion-nucleus scattering it is currently 
being applied to heavy-ion elastic scattering. These two 
methods will be treated shortly is some detail.
Charagi and Gupta \cite{cg} treated low-energy heavy-ion 
scattering in the Coulomb modified optical limit in two 
papers. Cha \cite{cha} introduced the deviation of the orbit in 
Glauber from the nuclear potential as well as the Coulomb 
potential. Charagi \cite{charagicomment} commented on the
Coulomb correction to the Glauber model introduced by Cha \cite{cha} 
saying that it failed to reproduce known reaction cross sections.
Alvi et al. \cite{alvi} calculated $\alpha-$nucleus scattering 
treating the system by phenomenology (fitting to Ni) and showed 
the Coulomb effect. Ahmad et al. \cite{ahmad} calculated 
$^{12}$C--$^{12}$C elastic scattering using the first two terms of an 
expansion in a Coulomb modified Glauber calculation.


\subsection{Organization}
The paper is organized as follows. In section \ref{mc} the method 
is described showing how to satisfy the center-of-mass condition
with the Monte Carlo sampling. In particular, the technique for
determining an auxiliary density which results in a desired center-of-mass
single-particle density is developed. In this section the method
for making the Coulomb correction is also presented.

In section \ref{results} the calculations are compared with data for
20 nuclear pairs/energies. The parameters for the auxiliary functions
are given. In section \ref{discussion} the results are summarized and
conclusions are drawn.
In the appendices we present a method for the rapid partial-wave 
projection of the Glauber amplitude and explain how the variation
calculation for the $^4$He density was done.

\section{Monte Carlo Eikonal Nucleus-Nucleus Scattering\label{mc}}

\subsection{The Glauber Method}

The eikonal theory of elastic scattering as expressed by 
Glauber \cite{glauber} and elaborated and studied by 
others \cite{sitenko,eikonal,harrington,amado,mcvoy} is believed to be an 
accurate representation of scattering at high energies and forward 
angles. It can be applied to calculate the multiple 
scattering of a simple projectile on a nucleus or collisions 
between nuclei, taking into account all of the scatterings 
possible in this theory.

There are two major difficulties in applying this technique: 
1) as the atomic number of the scatterers increases 
the number of scatterings becomes very large 
and 2) the representation 
of the wave function (density) of the two scattering 
conglomerates raises the problem of the center of mass.  The 
use of Monte Carlo techniques is well suited to solve both of these 
problems.

For the scattering of $\alpha$ particles on $^4$He (the 
largest case treated to date for the full scattering series 
{\em without} using Monte Carlo) there are 16 possible scatterings. 
Taking into account all of the possible orders there are 
$2^{16}-1=65535$ terms in the multiple scattering expansion. 
They are not all independent however and, by dividing them 
into classes, the calculation can be reduced \cite{yin} to 37 
different types, each characterized by a $4\times 4$ matrix 
of zeros and ones to be calculated and included with 
different weights. There remains a 24 dimensional integral to 
be done for each term. By using a Gaussian representation of 
the wave function of $^4$He, Franco and Yin \cite{fy} and 
Yin {\it et al.} \cite{yin} were able to provide expressions for 
these integrals and calculate the full sum.

In the present paper the Glauber expression for the multiple scattering 
is used directly as a product of factors in Eq. \ref{gtilde} without 
expanding into separate terms and the integral over the many-body 
nuclear density is done with Monte Carlo techniques, either with direct 
sampling or with the Metropolis algorithm. Normally one would be 
reluctant to use a Monte Carlo method to obtain a rapidly oscillating 
function as the amplitude for scattering, but in this case Monte Carlo 
can be used to calculate the profile function (which is a smooth 
function since it consists of a sum of analytic functions) and then a 
standard numerical technique can be applied to perform the last 
(one-dimensional) integral. The explicit development of the method is 
given in Section \ref{basic}.

The exact wave function for a nucleus at rest as a function 
of the coordinates of the nucleons (we are ignoring spin for 
the present) would necessarily be translationally invariant 
in the absence of external interaction. In order to calculate 
the scattering between the two nuclei one would simply 
translate the wave function to place the center of mass at 
the origin. Of course one rarely has the exact wave function 
available but it is possible to introduce an approximate 
density which is, indeed, translationally invariant where one 
can carry out this procedure. For $^4$He the four-body 
problem can be solved using Monte Carlo 
methods \cite{carlson,pudliner}. In the first instance a variational wave 
function can be used to find the minimum in energy by varying 
the parameters in the trial function. The trial wave function 
should be translationally invariant. The square of 
the wave function (the density, which is all that is needed 
for the Glauber scattering calculation) is represented by a 
collection of Metropolis walkers.

What has more commonly been done in practice is to start with 
an approximation to the density which is obtained empirically 
from a probe which is sensitive to the single-particle 
density relative to the center of mass, most often electron 
scattering.  One then applies some transformation to assure 
that the center of mass of the nucleus is at the origin. 
Franco and Yin \cite{fy} gave a prescription and showed that, 
using this formula, one could calculate with the fixed well 
assumption (for a harmonic oscillator potential) and then 
apply an exact correction. To illustrate the importance of 
this correction we point out that the factor by which the 
amplitude is multiplied for $\alpha-\alpha$ scattering at 
$-t=3$ (GeV/c)$^2$ is $3.6 \times 10^7$. Clearly, a careful 
treatment of the center-of-mass correlation is important. 
This problem is discussed at length in Section 
\ref{cmsection} and the MC method is applied the evaluation 
of the Glauber amplitude for $\alpha-\alpha$ scattering in 
Section \ref{aasection}.

It is important to define what one means by a measure of the 
center-of-mass correction, that is, we need to define the 
``zero effect'' condition. It is common in both the double 
folding model and the OLA to take the density from electron 
scattering (corrected for the charge distribution of the proton) to form a 
product density. Often the center-of-mass requirement is 
ignored in these calculations (see however Franco and 
Varma \cite{fv} for a first order treatment). We will take this 
product density as our ``no effect'' model and compare the 
scattering from this case with scattering from a density with the
proper center-of-mass properties. We will see that ignoring the 
center-of-mass effect, even for fairly heavy nuclei is ill advised.

\subsection{Basic technique\label{basic}}

The expressions for the Glauber amplitude are available many 
places, see e.g. Franco and Yin \cite{fy}. 

The nuclear profile 
function, $G(b)$ is the central function in the theory and is given by

\eq 
G(\bfb)=\int \prod_{i=1}^A d\bfs_i \prod_{j=1}^B d\bfs'_j
\phi^2(\{\bfs\}) \psi^2(\{\bfs'\})
G(\bfb,\{\bfs\},\{\bfs'\})\label{gtilde},
\qe
In Eq. \ref{gtilde} the notation $\{\bfs\}$ represents the collection of all 
of the coordinates of the projectile or target and 
$\phi(\{\bfs\})$ denotes the projectile ground-state wave 
function and $\psi(\{\bfs\})$ the target ground-state wave 
function.

\eq
G(\bfb,\{\bfs\},\{\bfs'\})\equiv 1-\prod_{i=1}^{A}\prod_{j=1}^B
\left[1-\Gamma_{ij}(\bfb+\bfs_i-\bfs_j)\right]. \label{basicpf}
\qe
The scattering amplitude is given by the two dimensional Fourier 
transform of $G(\bfb)$ and, since there is no direction defined for 
the nuclear wave function, it is just the Hankel transform of $G(b)$,
\eq
F(\bfq)
=\frac{ik}{2\pi}\int d^2b e^{i\bfq\cdot\bfb} G(\bfb)
=ik\int_0^{\infty} bdb J_0(qb) G(b).\label{ampl}
\qe
In Eq. \ref{basicpf} $\Gamma_{ij}(\bfb+\bfs_i-\bfs_j)$ denotes the 
two-dimensional Fourier transform of the elementary nucleon-nucleon 
amplitude $f(\bfq)$, i.e.,
\eq
\Gamma_{ij}(\bfb)=\frac{1}{2\pi ik}\int d^2q
e^{-i\bfq\cdot\bfb}f_{ij}(\bfq)=ge^{-b^2/2a}
\qe
$$
g=\sigma(1-i\rho)/(4\pi a)
$$
where the nucleon-nucleon amplitude has been approximated by
\eq
f_{ij}(q)=\frac{ik\sigma(1-i\rho)}{4\pi}e^{-\h aq^2};\ \ 
a=\ar+i\ai, \label{gausform}
\qe
where $\sigma$ is the nucleon-nucleon total cross section and $\rho$ is 
the ratio of the real to imaginary part of the forward amplitude. For a
discussion of the imaginary part of $a$, $a_I$ see Ref. \cite{phase}.

\begin{figure}[htb]
\epsfig{file=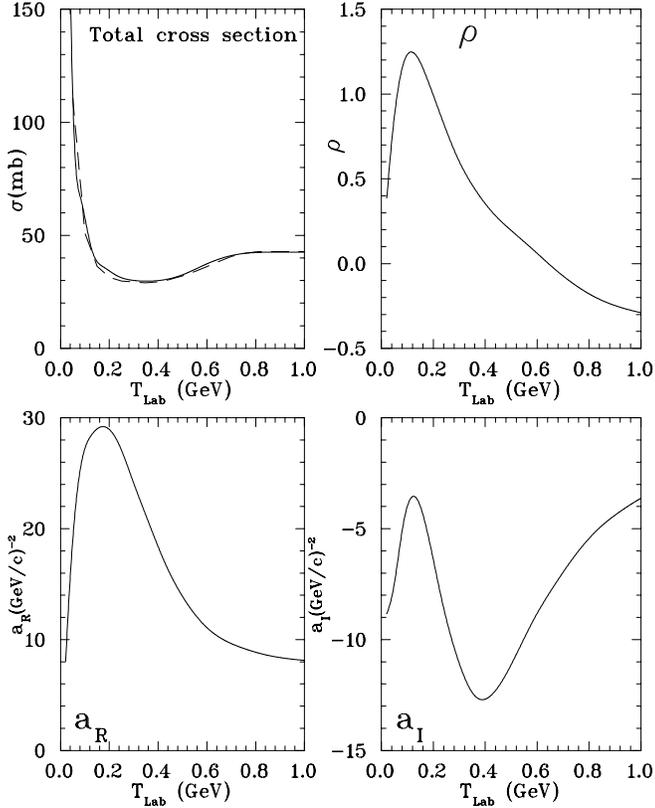,height=4.2in}
\caption{Nucleon-nucleon parameters taken from the work of Arndt et 
al. \cite{arndt}. Also shown for the average NN cross section is 
the recent parametrization of ref. \cite{bertulani} (dashed 
line). \label{parsarndt}}
\end{figure}

We have assumed a Gaussian approximation as a function of momentum 
transfer for the nucleon-nucleon amplitude. The scattering 
parameters needed can be extracted from nucleon-nucleon (NN) 
scattering data \cite{arndt} and such a set is shown in Fig. 
\ref{parsarndt} as a function of energy (see Refs \cite{wallace2} 
and \cite{pdg} for tables of other determinations of these 
amplitudes. Of course, one might expect that these amplitudes may 
well be modified in the nuclear medium (see e.g. Refs. 
\cite{sammarruca,furumoto,bertulani} but it may be useful to make 
calculations with free values to see how large the corrections are 
likely to be. We take the free values used in the calculations in 
this paper from the partial-wave parametrization of Arndt et al. 
\cite{arndt}. A plot of the free values used is shown in Fig. 
\ref{parsarndt}. For more details on the representation of the 
amplitudes, see the following section.

\begin{figure}[htb]
\epsfig{file=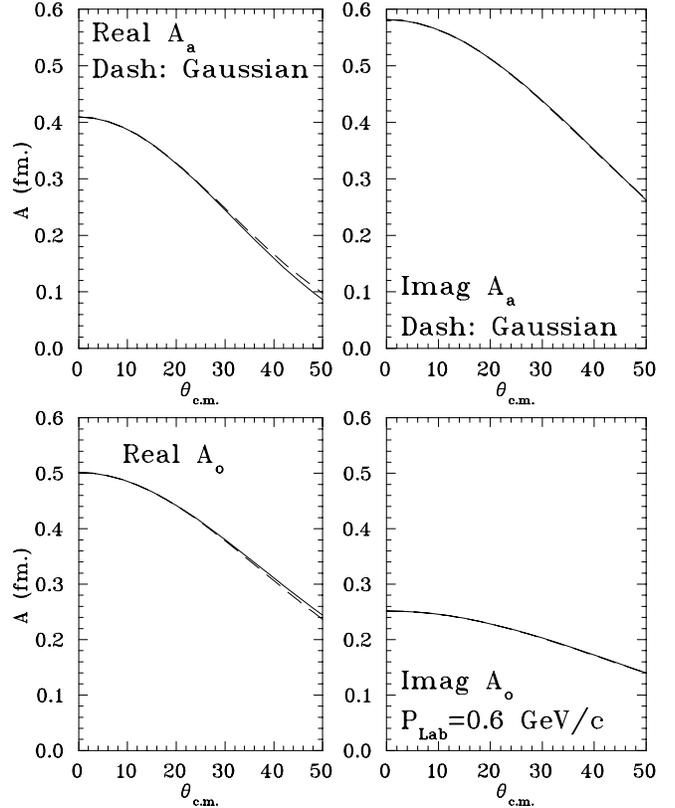,height=4.2in}
\caption{Angular distribution of the nucleon-nucleon amplitudes taken from the
phase-shift analysis of Arndt et al. \cite{arndt}. Also shown is the Gaussian 
approximation (dashed line) as used in Eq. \ref{gausform} with parameters chosen to 
match the value and slope in the forward direction. The ``aligned'' amplitudes
are shown in the upper panels and the ``opposite'' amplitudes in the lower panels.} 
\label{arndt2}
\end{figure}

\subsection{Nucleon-nucleon amplitude}

For NN free-space scattering there are essentially 4 incoherent 
beams in the free (and unpolarized) case according to the 4 
possible orientations of the spins of the two colliding 
nucleons. The imaginary parts of the forward elastic 
scattering amplitudes give the total cross section for each 
of these beams. The interference with the Coulomb in the 
forward direction can give the real parts of these amplitudes 
averaged with equal weighting.

This combined amplitude is given by
$$
A=\frac{1}{4}[M_{++++}+M_{----}+M_{+-+-}+M_{-+-+}]
$$
\eq
=\h[M_{++++}+M_{+-+-}] =\h[M_a+M_o]\label{amps}
\qe
with all amplitudes weighted equally. Here the amplitudes are 
labeled by ``aligned'' (subscript ``a'') and ``opposed'' 
(subscript ``o''). In the case of nucleus-nucleus scattering 
the amplitude would still consist of these two amplitudes if 
spin flip is neglected. Single spin flip amplitudes are very 
small in the forward direction, so are often neglected. 
Since we are considering the scattering of two spin-zero 
nuclei, a single spin flip is not possible and this small 
amplitude must enter twice in the calculation so that some 
other (correlated) nucleon can have its spin flipped in the 
opposite sense such that the total projection is again zero. 
These two constraints lead one to consider that the neglect 
of spin flip is a reasonable approximation.

However, even in the forward direction there is an amplitude 
$M_{+--+}$ which does not vanish at zero degrees so {\it a 
priori} might be expected to contribute. However, we see that 
this amplitude must flip the spin of one nucleon in each of 
the nuclei. The resulting spin projection must be reduced to 
zero again by a second spin flip in each nucleus. This 
amplitude arises primarily from one-pion exchange. The 
average over spin (and isospin) removes this amplitude from 
consideration, at least in first order and we neglect
it and consider only the amplitude coming from Eq. \ref{amps}. 

The total cross sections which correspond to the two terms in Eq. 
\ref{amps} have been studied experimentally and their difference 
shows a rapid variation \cite{madigan} sometimes attributed to a 
dibaryon (see e.g. \cite{mulders}). If some mechanism could be 
found to change the equal weighting of the aligned and opposite 
amplitudes in Eq. \ref{amps} then a change in the energy 
dependence of the nucleon parameters could be expected. However, 
the spin projections come from two different nuclei so it is 
difficult to see how such a correlation might come about. The 
amplitude in Eq. \ref{amps} does not correspond to any directly 
measurable differential cross section as a function of angle. Any 
attempt to extract the parameters directly from a measured 
differential cross section will lead to parameters not suitable 
for calculations using the Glauber expressions; the amplitude is a 
theoretical construct. Each of the terms in Eq. \ref{amps} can be 
well represented in the approximate form of Eq. \ref{gausform} 
(see Fig. \ref{arndt2}).

\begin{table*}[htb]
\begin{center}
\begin{tabular}{|ccccccccc|}
\hline
\hline
Case& T$_{Lab}$[Ref.]&T$_{Lab}$/A (MeV)&$\sigma$ 
(mb)&$\rho$&$a_R$&$a_I$&$
\theta_{Max}^{c.m.}(deg)$&-t$_{Max}$ [(GeV/c)$^2$]\\
\hline
$^6$He-$^{12}$C&0.230 
GeV\cite{lapoux}&38.3&163.5&0.680&14.6&-8.06&20&0.139\\
$^6$He-$^{12}$C&0,250 
GeV\cite{he6c1241.6}&41.6&159.7&0.737&16.0&-7.91&10&0.038\\
$\alpha$-$^{16}$O &0.240 GeV \cite{lui}&60&103.2&0.953&20.8&-6.65&12&0.051\\ 
$^{16}$O-$^{16}$O &1.120 
GeV \cite{nuoffer,khoa}&70&86.4&1.049&23.0&-5.87&22&1.224\\
$\alpha$-$^{208}$Pb& 0.288 GeV 
\cite{bonin}&72&83.0&1.069&29.2&-4.85&33&0.700\\ 
$\alpha$-$^{208}$Pb &0.340 GeV 
\cite{bonin}&85&65.0&1.169&25.7&-4.85&30&0.686\\ 
$^{12}$C-$^{12}$C &1.016 GeV 
\cite{buenerd}&85&65.0&1.169&25.7&-4.85&18&0.560\\
$^{16}$O-$^{40}$Ca &1.503 GeV \cite{oxygendat}&94&60.5&1.185&26.3&-4.56&6&0.254\\ 
$^{16}$O-$^{12}$C &1.503 GeV \cite{oxygendat}&94&60.5&1.185&26.3&-4.56&15&0.565\\ 
$\alpha$-$^{16}$O &0.400 GeV 
\cite{wakasa}&100&57.5&1.196&26.7&-4.36&37&0.800\\ 
$\alpha$-$^{208}$Pb &0.480 GeV \cite{bonin}&120&47.4&1.233&28.1
&-3.70&20&0.443\\ 
$^{12}$C-$^{12}$C &1.449 GeV \cite{hostachy}&120&47.4&1.233&28.1&-3.70&12&0.356\\
$^{12}$C-$^{12}$C &1.620 GeV \cite{ichihara}&135&43.1&1.212&28.6&-3.84&12&0.398\\
$\alpha$-$^{208}$Pb& 0.699 GeV \cite{bonin}&175&36.2&1.095&29.2&-5.05&12&0.240\\ 
$^{12}$C-$^{12}$C &2.400 GeV \cite{hostachy}&200&34.3&0.991&28.7&-6.42&10&0.410\\
$\alpha$-$^{40}$Ca& 1.37 GeV \cite{alphacadat}&343&29.7&0.435&20.5&-12.6&12&0.414\\ 
$\alpha$-$^{12}$C &1.37 GeV \cite{chaumeaux}&343&29.7&0.435&20.6&-12.6&21&0.799\\ 
$\alpha$-$\alpha$& 2.554
GeV \cite{berger}&638&38.9&0.010&10.5&-8.10&23&0.765\\
$\alpha$-$\alpha$& 4.20 
GeV \cite{satta}&1050&42.5&-0.303&8.06&-3.25&13&0.400\\ 
$\alpha$-$^{12}$C& 4.20 
GeV \cite{morschzp,morschprc}&1050&42.5&-0.303&8.06&-3.25&11&0.716\\ 
\hline
\hline
\end{tabular}
\end{center}
\caption{Parameters used in the calculations based on the nucleon-nucleon
partial-wave amplitude analysis of Ref. \cite{arndt}} \label{table1}
\end{table*}

Listed in Table \ref{table1} are values of the parameters used in the 
calculations presented in this paper. Also given are the maximum values of 
the center-of-mass angle and $-t$.

\begin{figure}[htb]
\epsfig{file=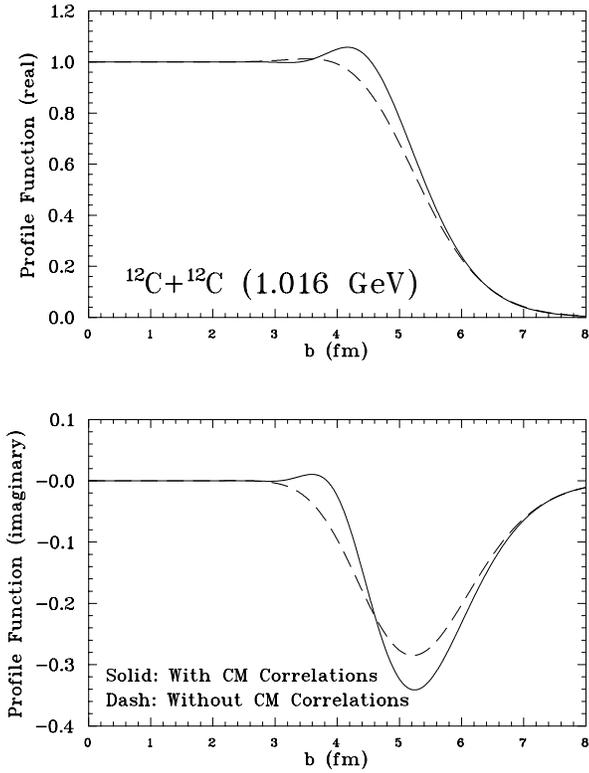,height=4.in}
\caption{The nuclear profile function for $^{12}$C--$^{12}$C 
scattering with (solid) and without (dashed) 
center-of-mass correlations. The details of the calculation 
are discussed in section \ref{ccsection}.} 
\label{bpool}
\end{figure}

\subsection{Nuclear profile function}

Another advantage of the method is that the nuclear profile 
function (NPF), 
$G(b)$, is available for study (see Fig. \ref{bpool}). This function
contains all of the information for the scattering since it is,
essentially, the Hankel transform of the amplitude (we are
dealing only with spin-zero on spin-zero scattering so there is
only one amplitude). Clearly the NPF is strongly dependent on the
single-particle density and the general shape reflects this. 

It is natural to ask how other characteristics of the nucleon
distribution, such as short-, medium- and long-range correlations.
They must manifest themselves in some manner but how? One can
get some feeling for this effect by considering the ``anatomy''
of the calculation. We see in Eq.~\ref{basicpf} that the NPF is expressed as one
minus a product. When this product goes to zero the NPF becomes
unity which corresponds to total absorption. When the impact
parameter, $b$, is large only one factor (at most) in the product
will differ from unity and only the single-particle density 
matters. As $b$ becomes smaller, more factors differ from
one and the product becomes smaller in magnitude.
Since the real part of the coefficient of the exponential in the 
function $\Gamma$, $g$, is less than unity over most of the range
of energies treated here each factor has modulus less than one.
The imaginary part, $a_I$, of $a$ is smaller than the real part 
$a_R$ and so plays a minor role in this qualitative discussion and 
we treat $a$ as real for that reason. 

\begin{figure}[htb] 
\epsfig{file=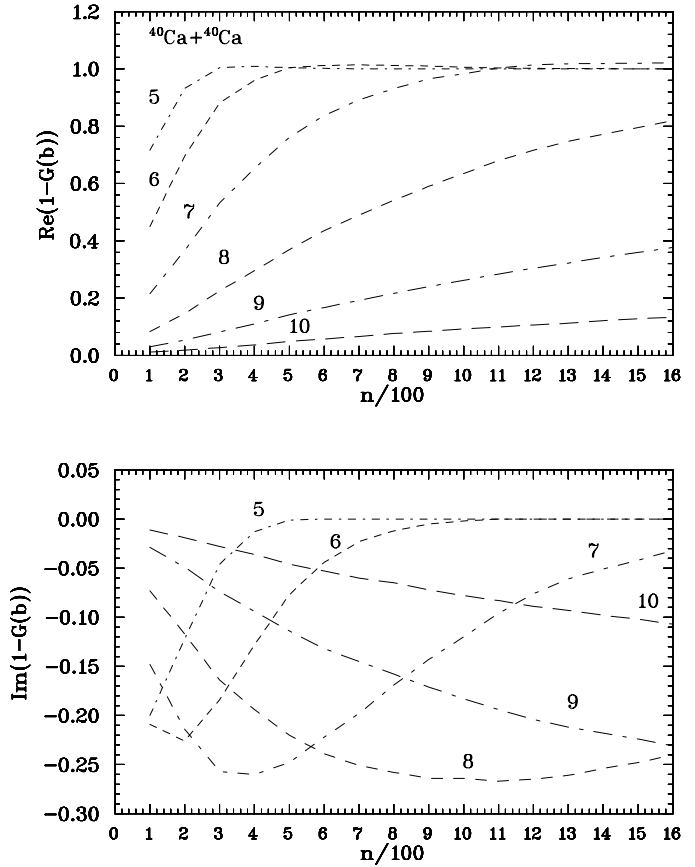,height=4.5in} 
\caption{One minus the accumulated nuclear profile function as a function
of the number of factors in the product in Eq. \ref{basicpf} for 
$^{40}$Ca-$^{40}$Ca scattering. The labels on the curves are the values of $b$ in 
fm.} 
\label{count} \end{figure}

Just how small each factor is depends on the distance 
$|\bfs_i-\bfs_j|$; if it is small then the factor is also small.
This difference does not depend on nucleon-nucleon correlations
since $\bfs_i$ and $\bfs_j$ are in different nuclei. However,
since ${\rm Real}\ g$ is of the order of 0.25-0.8, no one factor can
can drive the product to zero, it will take a combination of
several factors. This will require that several nucleons
be located close together and the probability of this occurring
is sensitive to the correlations.
Just how close together the nucleons have to be (in the 
2-dimensional space) is governed by $a_R$. Typical values
of $\sqrt{a_R}$ are in the range of 1 fm. Repulsive correlations
pushing the nucleons outside of this range would lead to greater
transparency. Correlations shorter than this range can be expected
to have little effect.

At the same time that the modulus of the product is 
decreasing it is developing a phase. Under the conditions 
just outlined the phase of each factor has the same sign so 
the phase grows monotonically as each factor is included.
Thus the behavior of the NPF depends on the relative sizes
of the phase and the modulus of the product. If the phase
reaches $\pi$ while the modulus of the product is still
sizable then the NPF will increase before becoming unity
as seen in Fig. \ref{bpool} with the CNC included. Further changes 
leading to the average modulus becoming larger (with only modification 
of the phase) could even lead to oscillations.

In an effort to get a feeling for the effective number factors (thus 
the number of times that the correction is being applied) we calculated 
the profile function for values of $b$ in the crucial range of the 
surface as a function of the number of factors (see Fig. \ref{count}). 
As we progress from large to small values of $b$ more factors are 
important. For large $b$ the profile function can be expressed as a sum 
of first-order terms and the dependence on the number of factors 
becomes linear. Of course in this limit there is no CMC since only 
first order in the density is involved.

\subsection{Center-of-mass treatment\label{cmsection}}

The scattering between two composite objects is treated in terms of the 
coordinate connecting their centers of mass so that the particles 
making up the nuclei must be centered appropriately. This is the 
problem of center of mass which must be appropriately addressed.
The Monte Carlo method can provide the solution.  Consider two methods 
of treating the c.m. motion. 
Each one provides a density in which the sum of all of the 
position vectors is zero but they are not (in general) equivalent.
The first method relies on Gaussian densities and provides
a correction factor to a scattering calculation made with an auxiliary
density centered about a fixed origin.

Franco and Yin \cite{fy} used this first method, developed by Czyz and 
Maximon \cite{cm} (see also Ref. \cite{gs}) which works only when using 
Gaussian densities because the center-of-mass can be expressed as a factor in 
this case. To implement this method, a model is made assuming 
a product of densities which are invariant under translation 
and using the algebraic identity:

$$
e^{-A\alpha^2R^2} 
e^{-\alpha^2(\bfr_1-\bfR)^2}e^{-\alpha^2(\bfr_2-\bfR)^2} 
e^{-\alpha^2(\bfr_3-\bfR)^2}\dots
$$
\eq
=e^{-\alpha^2r_1^2}e^{-\alpha^2r_2^2}e^{-\alpha^2r_3^2} 
\dots=\rho_a(r_1)\rho_a(r_2)\rho_a(r_3)\dots 
\qe
where $\bfR\equiv (\sum \bfr_i)/A$ is the center-of-mass 
coordinate and $\rho_a(r)$ is an auxiliary density with 
reference to a fixed origin. The expectation value in Eq. 
\ref{gtilde} can be taken over the auxiliary density (a 
calculation which is much easier) and the expectation value 
of the translationally invariant density obtained by dividing 
by the expectation value of the first factor on the left.

For the density used by Franco and Yin with $\rho_a(\bfs_i)\equiv 
\phi^2(\bfs_i)$
$$
|\chi(\{\bfs\})|^2=N
\delta\left(\frac{1}{A}\sum_j\bfs_j\right)\prod_{i=1}^A\rho_a(\bfs_i)
$$
\eq
=\frac{N}{(2\pi)^3}
\int d\bfQ\ e^{i\sum_j\bfQ\cdot\bfs_j/A}\prod_{i=1}^A\rho_a(\bfs_i)
\label{chidef}
\qe
where $N$ is a normalization factor. The single particle 
density relative to the center of mass, such as that obtained 
from electron scattering (see Refs. \cite{bertozzi,ottermann,gunten,
chandra} for corrections), for example, will be obtained, 
integrating over all but one of the coordinates, as
$$
\rho_s(\bfs_1)\equiv \int d\bfs_2 d\bfs_3 \dots |\chi(\{\bfs\})|^2
$$
\eq
= \frac{N}{(2\pi)^3}\int d\bfQ\ 
e^{i\frac{\bfQ\cdot\bfs_1}{A}}\rho_a(\bfs_1)\rho^{A-1}(\bfQ/A)
\qe
The Fourier transform  of the single particle density (in the c.m.)  
will be
$$
\rho_s(q)\equiv 
\int d\bfs_1\ e^{i\bfq\cdot\bfs_1}\rho_s(\bfs_1)
$$
$$
=\frac{N}{(2\pi)^3}\int d\bfQ\ \rho_a (\bfq+\bfQ/A)\rho_a^{A-1}(\bfQ/A)
$$
\eq
=\frac{A^3N}{(2\pi)^3}\int d\bfQ\ \rho_a(\bfq+\bfQ)\rho_a^{A-1}(\bfQ)
\qe
Normalizing, we have
\eq
\rho_s(q)=\frac{\int d\bfQ\ \rho_a (\bfq+\bfQ)\rho_a^{A-1}(\bfQ)}
{\int d\bfQ\ \rho_a^A(\bfQ)}
\qe
For a Gaussian form in momentum space as Franco and Yin 
\cite{fy} assumed
\eq
\rho_a(p)=e^{-p^2/4\alpha^2}
\qe
we can perform the integral on $\bfQ$ to find
{\large
\eq
\rho_s(q)=e^{-\frac{(A-1)q^2}{4A\alpha^2}}
\qe}
so that the auxiliary density has the same (Gaussian) form as the 
single-particle density but with a larger rms radius:\ $r_a^2=Ar_s^2/(A-1)$.

We can also calculate the transform of the full density (Eq. \ref{chidef}) 
in the general case as

$$
\rho_s(\{\bfq\})\equiv
\int d \bfs_1 d\bfs_2 \dots d\bfs_A\ e^{i(
\bfq_1\cdot\bfs_1+\bfq_2\cdot\bfs_2\dots +\bfq_A\cdot\bfs_A)}
|\chi(\{\bfs\})|^2
$$
$$
=
\frac{N}{(2\pi)^3}\int d\bfQ \prod_{i=1}^A \rho_a(\bfq_i+\bfQ/A)
$$
\eq
=\frac{\int d\bfQ \prod_{i=1}^A \rho_a(\bfq_i+\bfQ)}{\int d\bfQ\ 
\rho_a^A(\bfQ)}.
\qe

Evaluating with a Gaussian form we have
\eq
\rho_s(\{\bfq\})=e^{-\left[\sum q_i^2
-\frac{1}{A}(\sum \bfq_i)^2\right]/4\alpha^2}.
\qe

In the Monte Carlo method the algorithm leading to a density in the 
center-of-mass frame is to first select the A coordinates, $\bfu$, 
according to A independent auxiliary densities, $\eta_a(u)$, 
(assumed to be isotropic). The functions $\eta_a(u)$ have chosen 
forms and the principal aim of this section is to develop a method 
to pick the functions $\eta_a(u)$ such that they result in a 
specified single particle density relative to the center of mass. 
The coordinates to be used in the integration are obtained from the 
set of vectors $\bfu_i$ by

\eq
\bfs_i=\bfu_i-\frac{1}{A}\sum \bfu_j
\qe
which means that the Monte Carlo densities that result from this 
transformation produce (by construction) functions which have the 
sum of their vector coordinates zero.

To summarize, the procedure is to choose (isotropic) distributions 
according the density $\eta_a(u)$ for all of the nucleons in a 
given nuclear configuration. The center-of-mass vector is then 
computed and subtracted from each of the $\bfu_i$ to give the 
coordinates to be used in the evaluation of $G(\bfb,\{\bfs\},\{\bfs'\})$.

In order to choose an auxiliary function that gives a particular 
center-of-mass density it is very useful to have explicit 
expressions for the two functions that we need to connect.  The 
Fourier transform of the single particle density relative to the 
center of mass will be given by

$$
\eta_s(q)
=\int d\bfu_1d\bfu_2\dots d\bfu_A\ \times
$$
$$ 
e^{i\bfq\cdot(\bfu_1-\frac{1}{A}\sum \bfu_j)}\eta_a(u_1)\eta_a(u_2)
\dots \eta_a(u_A)=
$$
\eq
\eta_a\left(\frac{A-1}{A}q\right)\eta_a^{A-1}\left(\frac{q}{A}\right)
\label{relation}
\qe
To compare with the results of the previous section we can assume a Gaussian 
form again i.e. $ \eta_a(p)=e^{-p^2/4\beta^2} $ to get
\eq
\eta_s(q)=e^{-\frac{q^2}{4\beta^2}\frac{(A-1)^2}{A^2}}
e^{-\frac{q^2}{4\beta^2}\frac{A-1}{A^2}}
=e^{-\frac{A-1}{A}\frac{q^2}{4\beta^2}}
\qe
which (taking $\alpha=\beta$) gives the same result as the 
method used by Franco and Yin.
For the full Fourier transform we have
$$
\eta_s(\{\bfq\})
=\int d\bfu_1d\bfu_2\dots d\bfu_A\ \times 
$$
$$
e^{i\sum_{i=1}^A \bfq_i\cdot\left(\bfu_i
-\frac{1}{A}\sum_{j=1}^A\bfu_j\right)}
\eta_a(u_1)\eta_a(u_2)\dots \eta_a(u_A)
$$
$$
=\int d\bfu_1d\bfu_2\dots d\bfu_A\ \times
$$ 
$$
e^{i\sum_{i=1}^A \bfu_i\cdot\left(\bfq_i
-\frac{1}{A}\sum_{j=1}^A\bfq_j\right)}
\eta_a(u_1)\eta_a(u_2)\dots \eta_a(u_A)
$$
\eq
=\prod_{i=1}^A \eta_a\left(\bfq_i-\frac{1}{A}\sum_{j=1}^{A}\bfq_j\right)
\qe

If we take the Gaussian form again we have
$$
\eta_s(\{\bfq\})=e^{-\sum_{i=1}^A\left(\bfq_i-\frac{1}{A}
\sum_{j=1}^A \bfq_j\right)^2/4\beta^2}
$$
\eq
=e^{-\left[\sum_{i=1}^A\bfq_i^2-\frac{1}{A}
(\sum_{i=1}^A \bfq_i)^2\right]/4\beta^2}
\qe
which, with $\alpha=\beta$ again, is the same result as found previously. 
So we see that for Gaussian functions the two methods give identical results.

Eq. \ref{relation} is general assuming all nucleons have the same 
density. For the case where the number of protons and neutrons is not 
equal and the neutrons and protons have different density shapes a 
different form holds. If $\zeta_a(q)$ is the Fourier transform of the 
proton auxiliary function and $\zeta_s(q)$ is the Fourier transform of 
the proton single particle in the center of mass (and $\eta_a(q)$ and 
$\eta_s(a)$ are the corresponding functions for the neutrons) then the 
functions will be related by

\eq 
\zeta_s(q) =\zeta_a\left(\frac{A-1}{A}q\right)\zeta^{Z-1}_a(\frac{q}{A})
\eta_a^N\left(\frac{q}{A}\right)\label{proteq}
\qe
\eq 
\eta_s(q) 
=\eta_a\left(\frac{A-1}{A}q\right)\eta^{N-1}_a(\frac{q}{A})
\zeta_a^Z\left(\frac{q}{A}\right)\label{neuteq}
\qe

If the auxiliary functions for the protons and neutrons have rms radii $R_{pa}$ and
$R_{na}$ respectively the corresponding radii for the center-of-mass densities 
will be related by
\eq
R^2_{pc}=\left\{[(A-1)^2+(Z-1)]R^2_{pa}+NR^2_{na}\right\}/A^2
\qe
\eq
R^2_{nc}=\left\{[(A-1)^2+(N-1)]R^2_{na}+ZR^2_{pa}\right\}/A^2
\qe

For the case of a given form chosen for the single-particle 
density, we need to find an auxiliary density which satisfies Eq.
\ref{relation}
or
\eq
\eta_a(q)\ \eta_a^{A-1}\left(\frac{q}{A-1}\right)
=\eta_s\left(\frac{Aq}{A-1}\right).\label{relationetas}
\qe
One can find a solution to this equation for small $q$, by first 
expanding both $\eta_s(q)$ (assumed to be known) and $\eta_a(q)$ for small $q$.
\eq
\eta_a(q)=1-\mu q^2\dots; \ \ \ \eta_s(q)=1-\nu q^2\dots
\qe
Then we have to first order in $q^2$
$$
(1-\mu q^2)[1-\mu q^2/(A-1)^2]^{(A-1)}
$$
$$
=(1-\mu q^2)[1-\mu q^2/(A-1)]
$$
\eq
=
1-\mu q^2\frac{A}{A-1}=1-\nu q^2\frac{A^2}{(A-1)^2}
\qe

so that
\eq
\mu=\frac{\nu A}{A-1}
\qe
and the same relation holds between the rms radii in the general (Monte
Carlo) case as for the Gaussian case.

We could (in principle) solve Eq. \ref{relationetas} numerically on a mesh 
by constructing the first few elements on the mesh (3 of them) with the 
expansion. Then we can continue to calculate the remaining values on the 
mesh by evaluating the i$th$ point on the mesh with \eq 
\eta_a(q_i)=\eta_s\left(\frac{Aq_i}{A-1}\right)/\eta_a^{A-1}\left( 
\frac{q_i}{A-1}\right) \qe where the value of $\eta_a$ in the denominator 
is obtained by interpolation from previously calculated values on the mesh. 
These values would always be available since the argument is much smaller. 
This process works well for a Gaussian form but is not stable when there is 
a zero in the form factor.

In practice, we choose parametrized forms for $\eta_a(q)$ and vary the 
parameters using Eq. \ref{relation} to fit an (assumed known) function, 
$\eta_s(q)$. For convenience we choose functions to represent $\eta_a(r)$ 
which can be readily sampled directly (see e.g. Ref. \cite{book}).

The density with the center-of-mass effect is distinguished from the 
product density with the same single particle radial distribution by the 
fact that the distances between members of any pair of particles is greater 
than it would be for a simple product density. We can see this as follows.

For a single-particle density with independent particles the average of square of 
the distance between particles is given by
\eq
<(\bfr_1-\bfr_2)^2>=2<r^2>-2<\bfr_1\cdot\bfr_2>=2<r^2>\label{rnocor}
\qe
since the average over the independent particles gives a cross term
of zero. If we include a center-of-mass condition
\eq
\sum_{i=1}^A \bfr_i=0
\qe
then, replacing $\bfr_2$ in the cross product
$$
<\bfr_1\cdot\bfr_2>=-<r^2>-\sum_{i=3}^A<\bfr_1\cdot\bfr_i>
$$
\eq
=-<r^2>-(A-2)<\bfr_1\cdot\bfr_2>=-\frac{<r^2>}{A-1}
\qe
Thus, with center-of-mass correlations
\eq
<(\bfr_1-\bfr_2)^2>=2<r^2>\frac{A}{A-1}
\qe

For individual nuclear configurations we have
\eq
\frac{\sum_{i\ne j}(\bfr_i-\bfr_j)^2}{A(A-1)}
=\frac{2A}{A-1}\frac{\sum_{k}r_k^2}{A}
\qe

Thus we see that there is a correlation among pair of nucleons 
arising from the c.m. corrections. This condition is realized 
explicitly in the present method since the auxiliary density has a 
squared radius which is a factor of $A/(A-1)$ larger than the 
single particle density in the center of mass and, since the 
positions are independently chosen, Eq. \ref{rnocor} holds with the 
larger square radius. Since the displacement of the density to 
satisfy the center-of-mass condition does not change the distance 
between nucleons, this larger inter-particle distance is preserved 
while the distance of individual particles from the center of the 
nucleus is reduced.

With the procedure used here there is an initial density 
constructed in a Monte Carlo sense and then each realization is 
shifted by an amount to put the center of mass at the origin. In a 
spherical density with each particle thrown independently of the 
others, the rms distance between any pair of particles is 
$\sqrt{2}$ times the rms radius of the nucleus. Since the shift of 
the entire nucleus does not change the distance between pairs and 
the initial (auxiliary) density has a larger extent than the final 
density relative to the c.m. then the relative distance between 
pairs in the final density will be larger than the one which would 
be associated with a density constructed from independently thrown 
nucleons with the shape of the center-of-mass density. As the mass 
number, A, goes up this effect will become smaller with the 
relationship for the inter-particle radius squared being 
$R_{CMC}^2=A/(A-1) \times R_{No\ CMC}^2$. Hence one might assume 
that the center-of-mass effect goes to zero as A increases.

\begin{figure}
\epsfig{file=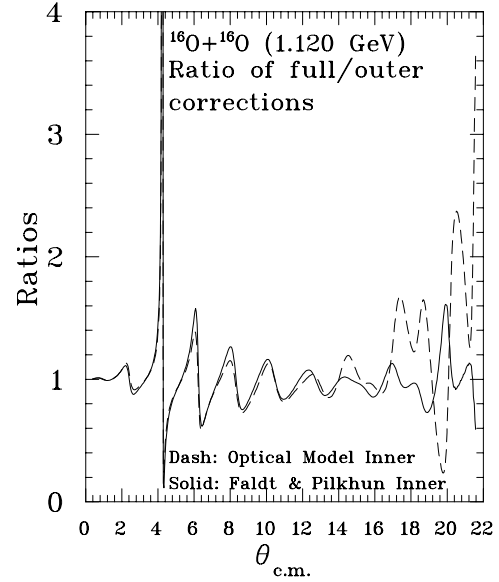,height=3in}
\caption{The ratio of inner+outer corrections to outer 
corrections only for the F\"aldt-Pilkuhn \cite{faldt} (point Coulomb) and
optical model with a realistic charge density. The full angular
distributions are given in section \ref{oosection}.}\label{inner}
\end{figure}

While it is true that the basic effect is becoming smaller, the 
calculation of a given observable may not. In the case that we are 
treating, of multiple scattering, the number of scatterings 
increases with A. Each scattering among the nucleons depends on the 
relative distance between nucleons and, while this distance is 
approaching that which would come from an independent particle 
density, the smaller effect is applied more times (there are more 
scatterings) so it is a numerical question as to whether the effect 
decreases or perhaps even increases with A. If the fundamental 
scattering interaction is weak so that only the low order 
scatterings are important then one can expect that the effect 
decreases with A. However, for the strong NN interaction it is 
possible that the importance of the CMC to multiple scattering does 
not decrease with A at all. Consider, for instance, the case of the 
scattering of a nucleus with A nucleons on its twin. In this case 
the number of scatterings goes as $A^2$ although there will be 
other factors which will limit the effective number of scatterings.

\subsection{Sampling considerations}

The multidimensional integral in Eq. \ref{gtilde} is to be 
done by Monte Carlo. The function to be averaged over, [G(b)], 
is complex but, as we shall see, the densities can be easily 
sampled. Ordinarily Monte Carlo is not very accurate if one 
takes a Fourier transform over a binned distribution (which 
we are not doing here).  The integral in Eq. \ref{ampl} is 
done by a standard quadrature method.

It is convenient to sample from pools of configurations of 
nuclei. For a variational or Monte Carlo Green's Function 
method for generating the nuclei this is the natural way to 
carry out the sampling process, but even if direct sampling 
is being done it is a useful method. One creates $N_t$ 
configurations for the target and $N_p$ for the projectile. 
Then M samples are taken in pairs, one from each pool, of all 
of the nucleons for a nucleus from the same configuration. If 
M is large enough the pairs will be identical for some cases. 
However, the number of identical pairs will be $M^2/N_tN_p$ 
for a fraction of repeated pairs of $M/N_tN_p$. Thus for a 
typical case of one million entries in each pool and 10 
million Monte Carlo samples the pairs repeat only 
$10^7/10^{12}=1/10^5$ fraction of the time.

This technique is also useful for calculating the effect of no 
center-of-mass correlations (CMC). One can simply take the 
coordinates for each nucleon from a {\it different} configuration. 
This guarantees that the same single-particle center-of-mass 
density is used but the coordinates in each nuclear configuration 
are uncorrelated.

The auxiliary functions are chosen with a form which can be
easily sampled. A common form used in many of the cases which
follow is
\eq
r^2\rho(r)=\sum g_n r^n\frac{b_n^{n+1}e^{-b_nr}}{n!}.
\qe
Often only two terms in the sum are needed to get an adequate
representation. To sample this function one term is
selected with probability, $g_n$, and then the corresponding
normalized probability function is sampled. The functions
in this case can be sampled easily by choosing $r$ by
\eq
r=-\frac{\ln \prod_{i=1}^{n+1} X_i}{b}
\qe
where the set, $X_i$, are independent random numbers uniformly
distributed between 0 and 1.

\subsection{Coulomb correction\label{coulsection}}

If the full amplitude with Coulomb is
$$
f(\theta)=\frac{1}{2ik}\sum_{\ell=0}^{\infty}(2\ell+1)(S_{\ell}-1)
P_{\ell}(\cos\theta)
$$
$$
=f_R(\theta)+\frac{1}{2ik}
\sum_{\ell=0}^{\infty}(2\ell+1)(S_{\ell}-e^{2i\sigma_{\ell}})
P_{\ell}(\cos\theta)
$$
\eq
=f_R(\theta)+
\frac{1}{2ik}\sum_{\ell=0}^{\infty}(2\ell+1)e^{2i\sigma_{\ell}}
(S_{\ell}e^{-2i\sigma_{\ell}}-1)P_{\ell}(\cos\theta)  
\qe
where $f_R(\theta)$ is the Rutherford amplitude and the 
$\sigma_{\ell}$ are the Coulomb phase shifts, and the amplitude 
without Coulomb is written
\eq
f_0(\theta)=\frac{1}{2ik}\sum_{\ell=0}^{\infty}(2\ell+1)
(S_{\ell}^0-1)P_{\ell}(\theta),
\qe
then we can write the factor between the two S-matrix elements as
\eq 
C_{\ell}=S_{\ell}e^{-2i\sigma_{\ell}}/S_{\ell}^0, \label{cdef}\qe
(see Ref. \cite{ga} and references therein). 
The inner Coulomb correction can be calculated with the use of an optical
model. The procedure is to calculate 
$C_{\ell}$ from Eq. \ref{cdef} using for $S_{\ell}$ the result of an
optical model fit with full Coulomb and for  $S_{\ell}^0$, the value
from the same optical model without the Coulomb interaction.
The quantity $C_{\ell}$ is then used as a correction factor with 
$S_{\ell}^0$ coming from a partial wave expansion of the Glauber
amplitude and $S_{\ell}$ being the Coulomb-corrected result.

Another way of getting the inner correction was given by F\"aldt and 
Pilkuhn \cite{faldt} as a simple shift in the value used in the integral 
over the profile function. 

\eq
F(\bfq)
=ik\int_0^{\infty} bdb J_0(qb) G(b')\label{ampfp}
\qe
where
\eq b'=\sqrt{b^2+\eta^2/k^2}+\eta/k, \qe
$\eta$ is the Coulomb parameter $\eta=ZZ'\alpha c/v$ and $k$
is the center-of-mass momentum. The Coulomb corrected S-matrix
element will be given by

\eq
S_{\ell}=e^{2i\delta_{\ell}}=
S^{FP}_{\ell}e^{+2i\sigma_{\ell}}
\qe
where $S^{FP}_{\ell}$ is the matrix element coming from the expansion
of Eq. \ref{ampfp} in partial waves.
For a similar method of treating the Coulomb
correction see Vitturi and Zardi \cite{vitturi}.
	
In order to carry out either of these corrections it is necessary to have the 
amplitude expressed as a partial wave sum. Since the eikonal calculation 
gives the amplitude as a function of momentum transfer or angle the 
projection of the amplitude onto partial waves is needed. The projection 
was not made directly but by using Eq. \ref{ampl} we can write

\eq 
S_{\ell}-1=-k^2\int_{-1}^1 dx \int_0^{\infty} b db P_{\ell}(x)J_0[q(x)b]G(b). 
\qe

The integral over $x$ of $P_{\ell}(x)J_0[q(x)b]$ can be carried out 
first very efficiently (see appendix A). The result was then 
integrated over $b$ with $b\ G(b)$. As a check the amplitude was 
then reconstructed from the S-matrix elements and compared with the 
original calculation. With the method given in Appendix A  
agreement was found within 0.1\%.

For the case of $^{16}$O-$^{16}$O scattering at 1.12 GeV we compared
the two methods. 
Figure \ref{inner} compares the two methods of making the inner correction 
for $^{16}$O+$^{16}$O (see Sec. \ref{oosection} for details). It displays 
the ratio of the fully corrected amplitude to that with only the outer 
correction. It is remarkable how well the two methods agree. The 
disagreement beyond 16 degrees is due mainly to the fact that the optical 
model is calculated with a finite charge distribution and the method of 
F\"aldt and Pilkuhn assumes a point charge distribution. For other 
treatments of the Coulomb correction, see Refs \cite{bertulanihansen} and 
\cite{bertulanigade}.

\section {Application of the Monte Carlo method: 
results\label{results}}

In this section we carry out calculations using the full Glauber formalism 
with values of the parameters taken from fits to the nucleon-nucleon 
amplitudes \cite{arndt}. In some cases we will vary those parameters to 
investigate the sensitivity to their values. Table \ref{table1} summarizes 
the cases treated, the data sources and the free-space parameters used.

\subsection{$\alpha-\alpha$ scattering \label{aasection}}
\begin{figure}[htb]
\epsfig{file=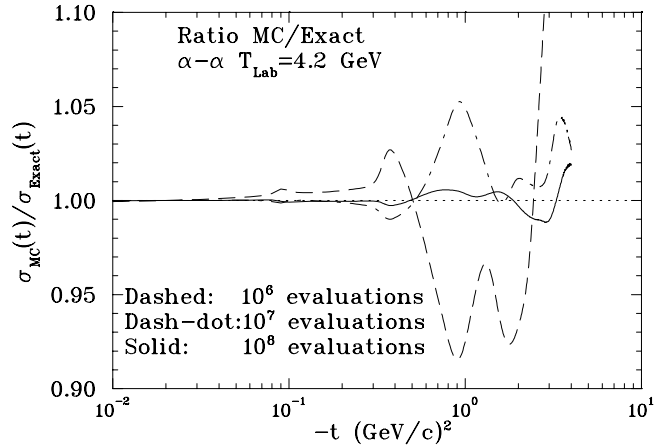,angle=90,height=2.3in}
\caption{Ratio of calculations for $\alpha-\alpha$ scattering for 
different numbers of Monte Carlo evaluations showing the degree of 
calculational precision obtained. The $\alpha$-particle density used
is the same one used by Franco and Yin \cite{fy}} 
\label{g44ratio}
\end{figure}

There are some interesting points for the $\alpha-\alpha$ scattering 
calculation. Notice that there is no separate center-of-mass 
correction factor as there was in the formulation of Franco and Yin 
\cite{fy}.  The Monte Carlo method calculates directly with the 
many-body density with the method outlined above.  As a test of the 
method we repeat the calculation of Franco and Yin \cite{fy} using 
their technique and the Monte Carlo method using Gaussian densities. 
From the previous discussion on the center-of-mass correction we 
should expect to find the same result aside from Monte Carlo 
statistical errors. Figure \ref{g44ratio} shows the ratios of the 
calculations done in the two ways. It is seen that the results are the 
same to about 2\% even though the absolute magnitude of the cross 
section changes by almost 10 orders of magnitude over the angular 
range considered. The difference between the two calculations is 
invisible on a log scale. For $10^8$ Monte Carlo evaluations the 
calculation took ~10 hours on a standard 2.5 GHz PC.

\begin{figure}[htb]
\epsfig{file=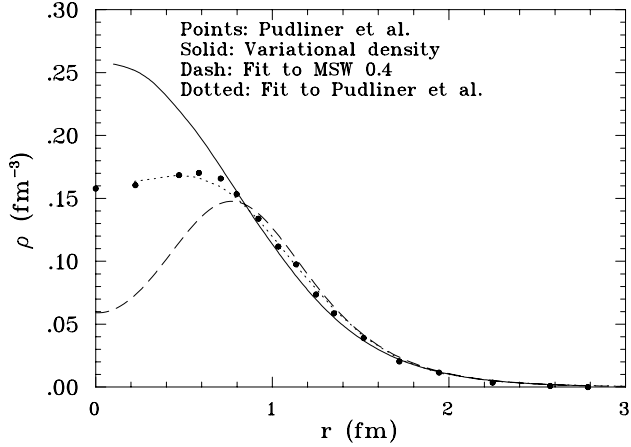,angle=90,height=2.3in}
\caption{Comparison of three densities used for the $\alpha$ 
particle in the calculations. The reference MSW is to McCarthy, 
Sick and Whitney \cite{msw}. The points were read from 
Fig. 15 in Pudliner et al. \cite{pudliner}}\label{alphadens}
\end{figure}

The nucleon density of $^4$He has been somewhat of a puzzle. The fits to 
electron scattering \cite{msw} with the method of a sum of Gaussians (SOG) 
show a large depression in the center of the nucleus that the Monte Carlo 
Green's Function calculations do not find \cite{pudliner}. However, the 
$\chi^2$ per data point is considerably less than unity showing that perhaps 
an over-fit was present in the SOG fit. Using other forms to represent the 
density, acceptable fits with a significantly smaller depression can be found. 
We have made a fit to the electron-scattering data \cite{msw} with a density 
in which we constrained the ratio of the density at the origin to that at the 
peak to be 0.4 in order to have a density similar to that found in Ref. 
\cite{msw} but with a less severe depression. That density is shown in Fig. 
\ref{alphadens} as the dashed line. We fit this density to find an auxiliary 
density given by

$$
r^2\rho_{\rm 
MSW}(r)=(1-\alpha)N_1\left(e^{-a_er}-e^{-b_er}\right)^{10}
$$
\eq
+\alpha[0.95N_2e^{-(r-r_0)^2/a_0^2}
+0.05N_3e^{-(r-r_1)^2/a_1^2}] 
\qe
where $\alpha=0.43$ and
\vspace*{-.1in} 
$$
a_e=0.51 {\rm fm}^{-1};\ b_e=0.61 {\rm fm}^{-1};
$$
\vspace*{-.3in} 
$$
 r_0=1.28\ {\rm fm};\ r_1=1.8\ {\rm fm};
$$
\vspace*{-.1in} 
\eq 
a_0=0.1\ {\rm fm}; \ a_1=0.3\ {\rm fm}
\qe
$N_1, N_2$ and $N_3$ are chosen to normalize each of the individual 
probability densities. The calculations of Pudliner et 
al. \cite{pudliner} give perhaps the best estimate of the $^4$He 
density and show only a slight depression in the central region.  
Also shown in Fig. \ref{alphadens} is a fit to the Monte Carlo 
density of Pudliner et al. \cite{pudliner} (large dots) from reading the points 
from the graph. The auxiliary density for this fit is given by
\eq 
r^2\rho_{\rm Pud}=\lambda_4 d^{15}r^{14}e^{-dr}/14!+
(1-\lambda_4)f^9r^8e^{-fr}/8! 
\qe 
where $\lambda_4=0.835$ and the resulting density is shown in 
Fig. \ref{alphadens} (dotted line). In order to have a density with short range 
correlations included we have carried out our own variational 
calculation (See Appendix B).

\begin{figure}[htb]
\epsfig{file=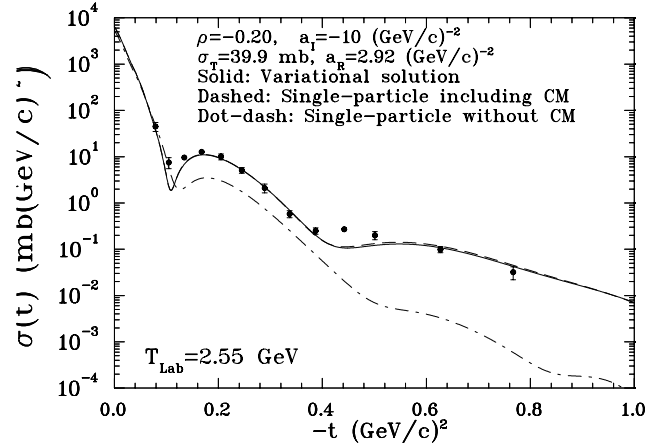,angle=90,height=2.3in}
\caption{Calculations of $\alpha-\alpha$ scattering showing the effect 
the center-of-mass correlations. The dash-dot curve uses the same 
Metropolis density as the solid curve but each nucleon is drawn from a 
different (random) realization of the nucleus. Thus it has the same 
single particle density but the effect of the CMC is missing. 
The data are from Berger et al. \cite{berger} }
\label{m4420m1mnc}
\end{figure}

To test for the relative importance of the short-range 
correlations and the center-of-mass correlations, we performed the 
calculation as outlined in the previous section starting from the 
single-particle density resulting from a density derived from a 
binning of the variational result. The c.m. density from that 
calculation is also shown in Fig. \ref{alphadens} (solid line).

\begin{figure}[htb]
\epsfig{file=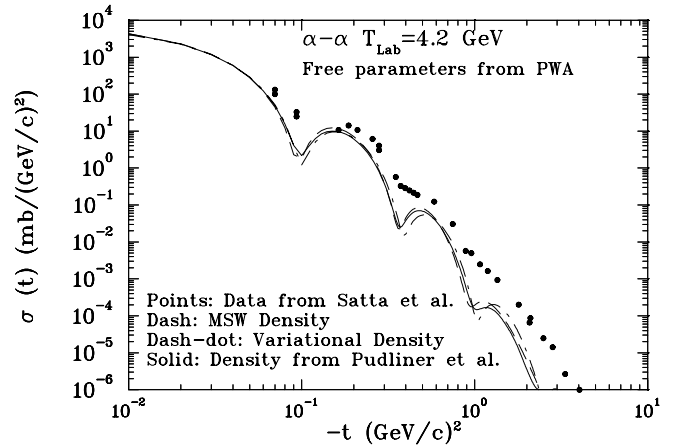,angle=90,height=2.3in}
\caption{$\alpha-\alpha$ scattering at T$_{\rm Lab}=4.2$ GeV showing the 
dependence on the density used. The notation MSW refers to the fit to 
the electron scattering fit by McCarty, Sick and Whitney \cite{msw} as 
described in the text. The partial wave analysis (PWA) is that of
Ref. \cite{arndt}. The data are from Satta et al. 
\cite{satta}\label{alphaalpha7gev/c}}
\end{figure}

The auxiliary density, $\eta_a(r)$, as fit to the variational density 
can be expressed as
\eq
r^2\eta_{\rm Var}(r)=(1-\alpha)g_1(r)+\alpha g_2(r)
\label{hefitform}\qe
where
\vspace*{-.1in}
$$
g_1(r)=N_g r^2e^{-a_gr^2};\ \ g_2=N_e(e^{-a_er}-e^{-b_er})^2
$$
\vspace*{-.3in}
$$
\alpha=0.505,\  a_g=0.56\ {\rm fm}^{-2},
$$
\vspace*{-.3in}
\eq
 a_e=0.779\ {\rm fm}^{-1},\  b_e=1.345\ {\rm fm}^{-1}
\qe
where $N_g$ and $N_e$ were chosen to normalize $g_1$ and $g_2$ 
individually. Note that the volume element, $r^2$, is included
in each case. 
Since the wave function has an asymptotic limit of the form 
$e^{-\kappa r}/r$ then, with the volume element included the tail 
should behave as an exponential.

\begin{figure}[htb]
\epsfig{file=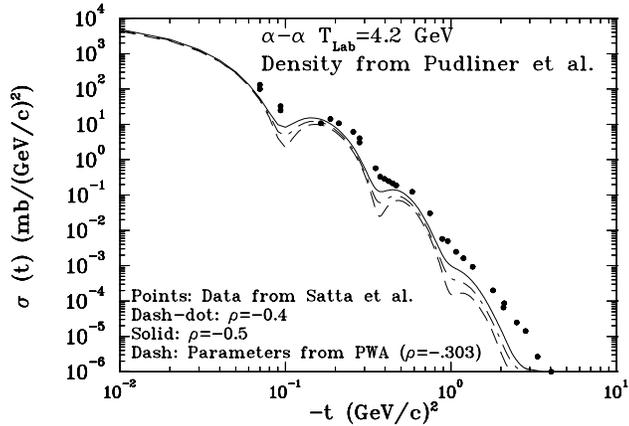,angle=90,height=2.2in}
\caption{$\alpha-\alpha$ scattering at 4.2 GeV showing the variation
with the parameter $\rho$. The data are from Satta et al. \cite{satta}.
PWA means the partial wave analysis from Ref. \cite{arndt}
 \label{alphaalpha7gev/crho}}
\end{figure}

\begin{figure}[htb]
\epsfig{file=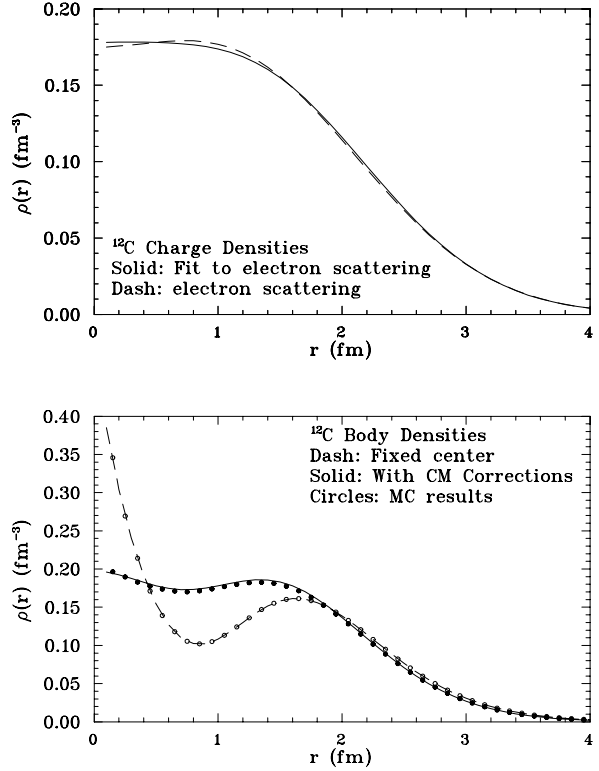,height=4.in}
\caption{The top panel shows the fit to the $^{12}$C charge 
density using the auxiliary density in Eq. \ref{auxc12}. 
The electron-scattering data are from Ref \cite{msc12}.
lower panel shows the point density along with the auxiliary 
density. Also shown in the lower panel are the points 
obtained from binning the radius values in the Monte Carlo 
calculation, the open circles corresponding to the auxiliary 
density and the solid circle to the center-of-mass density.}
\label{c12den}
\end{figure}

\begin{figure}[htb]
\epsfig{file=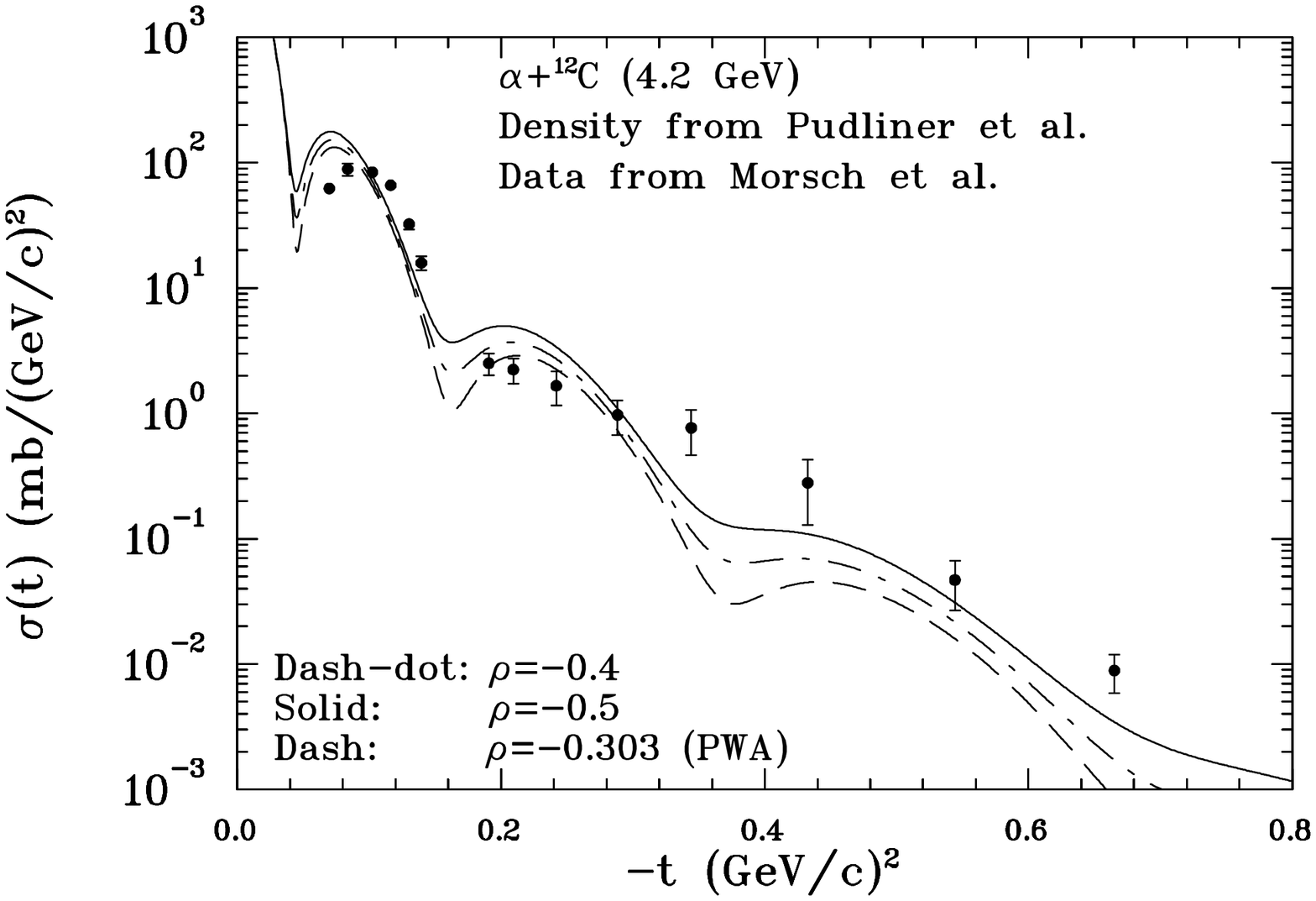,angle=90,height=2.2in}
\caption{$\alpha-^{12}$C scattering at 4.2 GeV showing the variation
with the parameter $\rho$. The data are from Morsch et 
al. \cite{morschzp,morschprc}. PWA means the partial wave 
analysis from Ref. \cite{arndt}.\label{alphac7gevcrho}}
\end{figure}

\begin{figure}[htb]
\epsfig{file=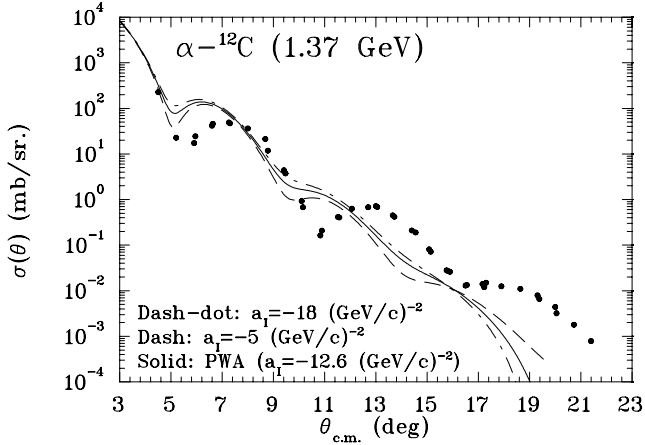,angle=90,height=2.3in}
\caption{Comparison of $\alpha$-$^{12}$C scattering with 
data of Chaumeaux et al. \cite{chaumeaux} showing the effect of
a variation of the nucleon-nucleon parameter $a_I$. PWA refers to the partial wave
analysis of Ref. \cite{arndt}. The density for the $\alpha$ projectile is
taken from Pudliner et al. \cite{pudliner} as described in the text.} 
\label{vpar1.37ai}
\end{figure}

Figure \ref{m4420m1mnc} shows a test of the importance of 
short-range correlations and center-of-mass correlations. For 
this illustration only, the parameters have been chosen to give a 
reasonable representation of the data, unlike most other figures 
presented which use the free parameters determined from amplitude 
analyses \cite{arndt}. The solid curve shows the calculation with 
the full variational wave function and the dashed curve shows the 
results obtained with the auxiliary density used to correct for the 
center of mass. The dash-dot curve shows the result of choosing 
each nucleon from a different configuration of the nucleus. In this 
case the single particle density will be identical to the other two 
cases but there is no CMC among the nucleons. The principal 
difference between the two is that the solid curve does not have 
the short-range correlations that are in the variational wave 
function. Hence, at least for the variational calculation performed 
here (see Appendix B), they are not important in agreement with Ullo
and Feshbach \cite{ullo}.

Figure \ref{alphaalpha7gev/c} shows the dependence on the density
used. It is seen that the sensitivity is very small and reasonable
variations would not seem to be able to significantly improve the agreement
with the data.

Figure \ref{alphaalpha7gev/crho} shows a comparison of the calculation 
at 4.2 GeV beam energy with the data of Satta et al. \cite{satta}. 
Also shown are the results of a variation with the $\rho$ 
parameter. This parameter is the least well determined 
experimentally of the nucleon-nucleon parameters. It is seen that 
the prediction is rather poor when compared with the data and that 
moderate variations of $\rho$ are unlikely to improve the 
agreement.

\subsection{$\alpha$-$^{12}$C scattering\label{alcsection}}

\begin{figure}[htb]
\epsfig{file=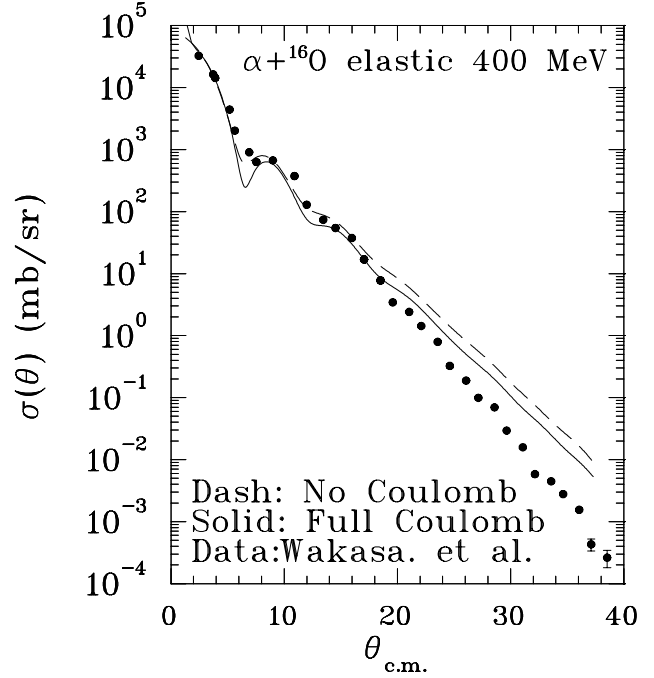,height=3.5in}
\caption{Comparison of $\alpha$-$^{16}$O scattering with 
data of Wakasa et al. \cite{wakasa} showing the effect of the Coulomb
correction.} 
\label{alox400}
\end{figure}

The carbon auxiliary density was obtained from the modified 
harmonic oscillator fit of charge density obtained from electron 
scattering data \cite{msc12}. The form found for the auxiliary 
density is
\eq
r^2\eta_{12}(r)=0.90916\ h_1(r)+0.09084\ h_2(r)\label{auxc12}
\qe
where
\eq 
h_1(r)=M_1r^{10}e^{-d_{12}r};\ \ h_2(r)=M_2(e^{-a_{12}r}-e^{-b_{12}r})^2
\qe
and $d_{12}=0.7311$ fm$^{-1}$,\ $a_{12}=1.7488$ fm$^{-1}$;\ 
$b_{12}=0.6429$  fm$^{-1}$.
$M_1$ and $M_2$ are chosen to normalize $h_1$ and $h_2$ separately.

\begin{figure}[htb]
\epsfig{file=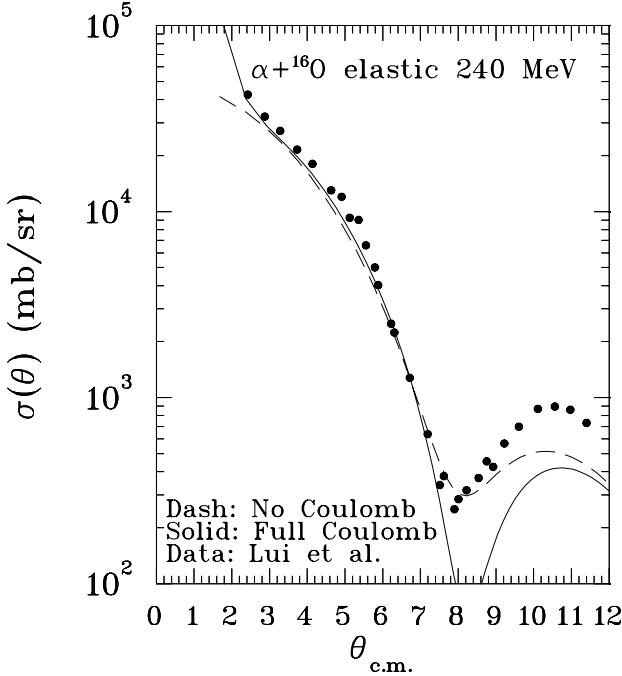,height=3.5in}
\caption{$\alpha-^{16}$O scattering at 240 MeV.
 The data are from Lui et al. \cite{lui} \label{alox240}}
\end{figure}

Figure \ref{c12den} shows the auxiliary density and the 
center-of-mass body and charge densities which result. For the 
light nuclei the auxiliary density is normally quite different from 
the center-of-mass density.


Figure \ref{alphac7gevcrho}  shows the variation of the differential cross 
section at 4.2 GeV with the parameter $\rho$. As for  $\alpha$-$\alpha$ 
scattering at the same energy (Fig. \ref{alphaalpha7gev/c})  the moderate 
variations of $\rho$ do not really improve the agreement with the data. 
Figure  \ref{vpar1.37ai}  shows the prediction of the calculation for 
$\alpha$-$^{12}$C scattering at 1.37 GeV kinetic energy with the parameter 
$a_I$. The agreement is no better than that at 4.2 GeV and variations of 
the other parameters give similar results. Again, it seems unlikely that a 
modest variation in parameters will bring the calculation in line with the 
data.

\begin{figure}[htb]
\epsfig{file=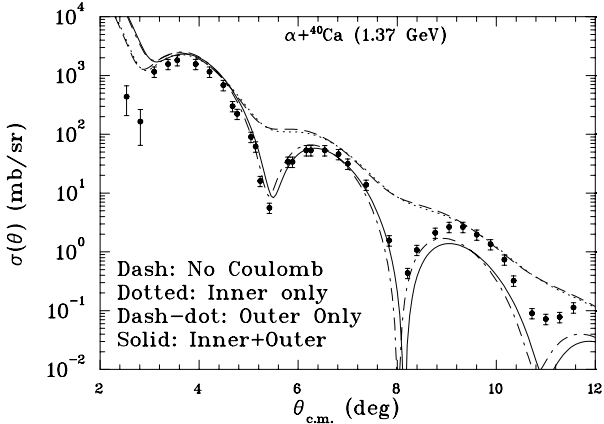,angle=90,height=2.2in}
\caption{$\alpha$-$^{40}$C scattering at 1.37 GeV compared with the 
data of Alkhazov et al. \cite{alphacadat}}
\label{alphaca}
\end{figure}

\begin{figure*}[htb]
\epsfig{file=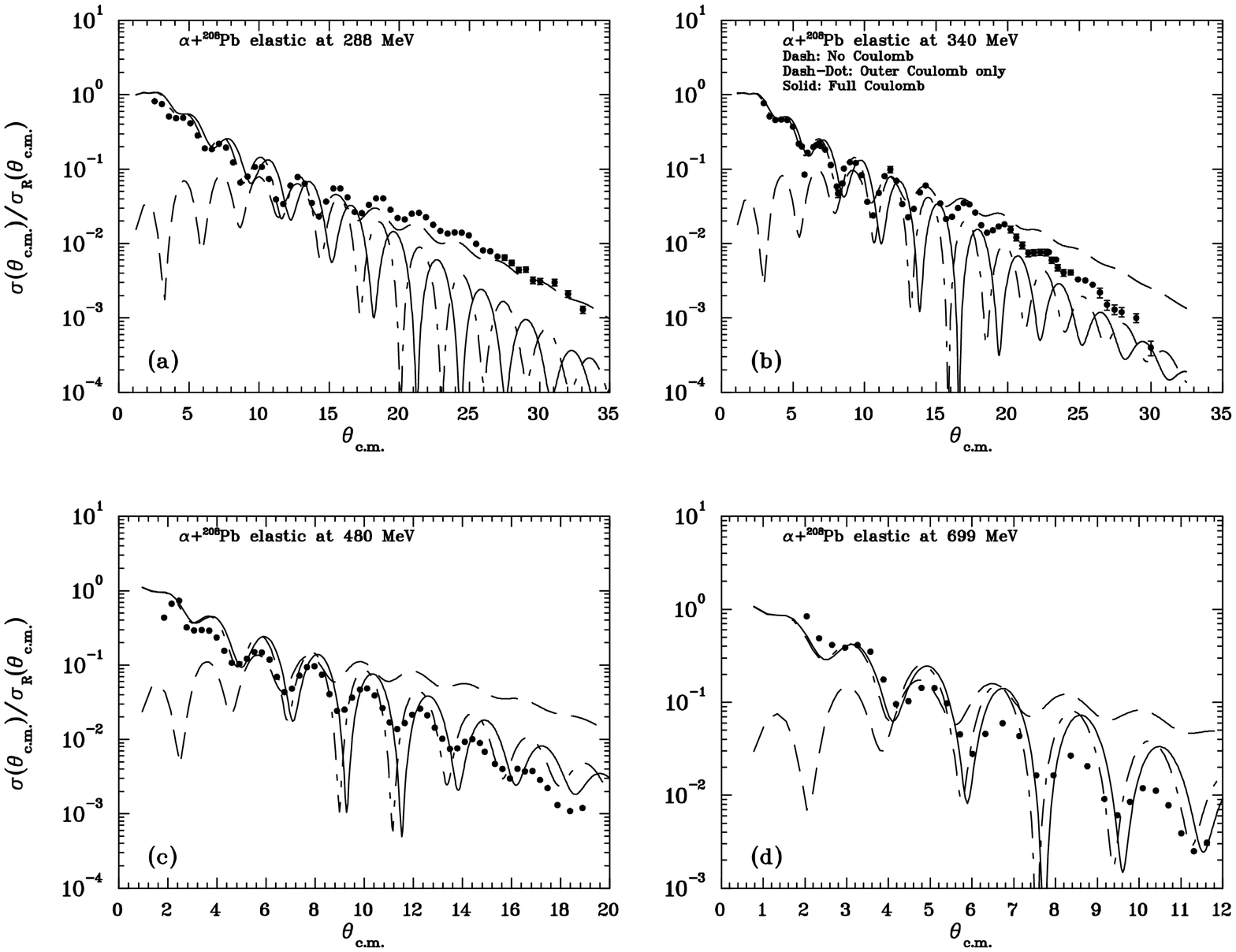,angle=90,height=5.5in}
\caption{$\alpha$-$^{208}$Pb scattering compared with the
data of Bonin et al. \cite{bonin}}
\label{alpb}
\end{figure*}

\subsection{$\alpha$-$^{16}$O scattering\label{aloxsection}}

The auxiliary function for $^{16}$O was taken to have the form:
\eq r^2\eta_{16}(r)=0.8646\  v_1(r)+0.1354\ v_2(r)
\qe
with
\eq v_1(r)=M_1r^{10}e^{-d_{16}r};\ \  
v_2(r)=M_2(e^{-a_{16}r}-e^{-b_{16}r})^2
\qe
and
\eq
a_{16}=2.738\ {\rm fm}^{-1};\ b_{16}=0.2976\ {\rm fm}^{-1};\ 
d_{16}=4.174\ {\rm fm}^{-1}
\qe
with 
$$M_1=d_{16}^{11}/10!$$
and 
$$M_2=2a_{16}b_{16}(a_{16}+b_{16})/(a_{16}-b_{16})^2$$ 
chosen to normalize $v_1$ and $v_2$.

Figure \ref{alox400} shows the results of the calculation 
$\alpha$-$^{16}$O compared with the data of Wakasa et 
al. \cite{wakasa} at 400 MeV. Here the agreement at forward angles 
is satisfactory. Figure \ref{alox240} compares the calculation at 
240 MeV with the data of Lui et al. \cite{lui} here the agreement is 
fairly good (aside from a slight normalization problem) up to about 
7 degrees. After that there is a considerable difference. The 
Coulomb correction is substantial, especially at larger angles.

\subsection{$\alpha$-$^{40}$Ca scattering\label{alcasection}}

The auxiliary density for Ca was obtained from a point density 
extracted from pion scattering \cite{gd} which is very similar to 
that obtained from electron scattering \cite{sickca40}. The form 
is
\eq r^2\eta_{40}(r)=0.740072\  v_1(r)+0.259928\  v_2(r)
\qe
with
\eq v_1(r)=M_1r^{10}e^{-d_{40}r};\ \  
v_2(r)=M_2(e^{-a_{40}r}-e^{-b_{40}r})^2
\qe
and
\eq
a_{40}=0.641\ {\rm fm}^{-1};\ b_{40}=0.443\ {\rm fm}^{-1};\ 
d_{40}=3.597\ {\rm fm}^{-1}
\qe
with $M_1$ and $M_2$ chosen to normalize $v_1$ and $v_2$.

The results of a calculation for $\alpha$-$^{40}$Ca scattering at 
1.37 GeV are shown in Fig. \ref{alphaca} and compared to the data 
of Alkhazov et al. \cite{alphacadat}. The Coulomb correction plays a 
very large role. There would be no agreement at all without it. 
While the agreement is far from perfect, it is satisfactory for 
no adjustable parameters at least for the angles less than 8 
degrees (with the exception of the two forward data point which 
have large errors). For other treatments of $\alpha-^{40}$Ca scattering
see Refs. \cite{hassanmetawei} and \cite{ahmadalpha}.

\subsection{$\alpha$-$^{208}$Pb Scattering \label{alpbsection}}

The auxiliary density for $^{208}$Pb was fit to the charge density of 
$^{208}$Pb so corresponds to the proton density. It is believed that the 
neutron density is different from the proton density but a study of the 
effect of a different neutron density, interesting though it might be, is 
beyond the scope of the present work. Since the lead nucleus is much larger 
than the other nuclei considered so far, with a significant flat portion 
inside the surface, the forms of the auxiliary density used for the lighter 
nuclei are not appropriate for this case. Because of the large number of 
nucleons, the difference between the auxiliary density and the center-of-mass 
density is not very great. For these reasons a Woods-Saxon (WS) density was 
used, fit to the charge density \cite{devries}. The volume element must be 
included as well in the sampling as has been done in previous cases. While 
the WS density can be directly sampled, it is the form

\eq
\frac{r^2}{1+e^(\frac{r-c}{a})}
\qe
which is needed. This sampling was done by the method of selection of 
variables \cite{book}. A comparison was made between variables from two samples. First, 
$r_1$ is chosen according to a linear distribution from 0 to $R_0$ then a second 
value, $r_2$ was obtained from the WS distribution with 
$$
r_2=a\ln\left\{\frac{1}{b[(1+1/b)^F-1]}\right\}
$$
where $F$ is a random number uniformly distributed between 0 and 1 and 
$b=e^{-c/a}$. If $r_2$ is greater than $r_1$ it returned as the desired value of 
$r$. Otherwise the process is repeated. The $\alpha$ particle density used was 
the one 
fit to Pudliner et al. \cite{pudliner} as discussed earlier.

Figure \ref{alpb} shows the results of calculations at energies of 288, 340, 480
and 699 MeV compared with the data of Bonin et al. \cite{bonin}. The Coulomb
correction plays a very large role in determining the cross section as might
be expected. The agreement is rather good in the forward direction but worsens
rapidly beyond a certain angle, different for each energy.

\begin{figure}[htb]
\epsfig{file=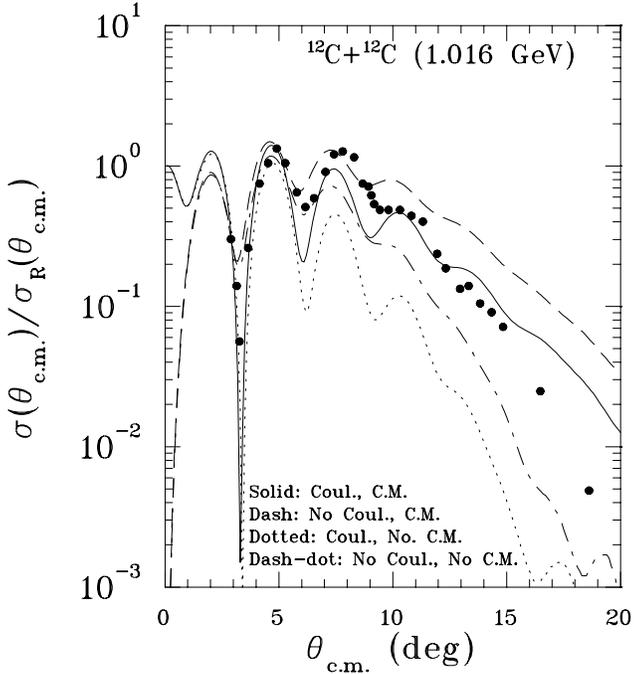,height=3.5in}
\caption{$^{12}$C$-^{12}$C scattering at 1.016 GeV with and 
without the center-of-mass correction. The data are from Buenerd
et al. \cite{buenerd} } 
\label{cc1.016cm}
\end{figure}

\subsection{$^{12}$C-$^{12}$C Scattering\label{ccsection}}

A comparison with the data at 1.016 GeV \cite{buenerd} and 1.45 and 2.4 
GeV \cite{hostachy} has been been made by several groups \cite{hassanain, 
abu4,abu3,lenzi1,lenzi2,elg2,elg1,mansourmetawei,ahmad,chauhan,abu} 
using various methods. See also the phase-shift analysis by Mermaz et al.
\cite{mermaz}. We can use the $^{12}$C densities found in section 
\ref{alcsection} to calculate the scattering of $^{12}$C from $^{12}$C.

\begin{figure*}[htb]
\epsfig{file=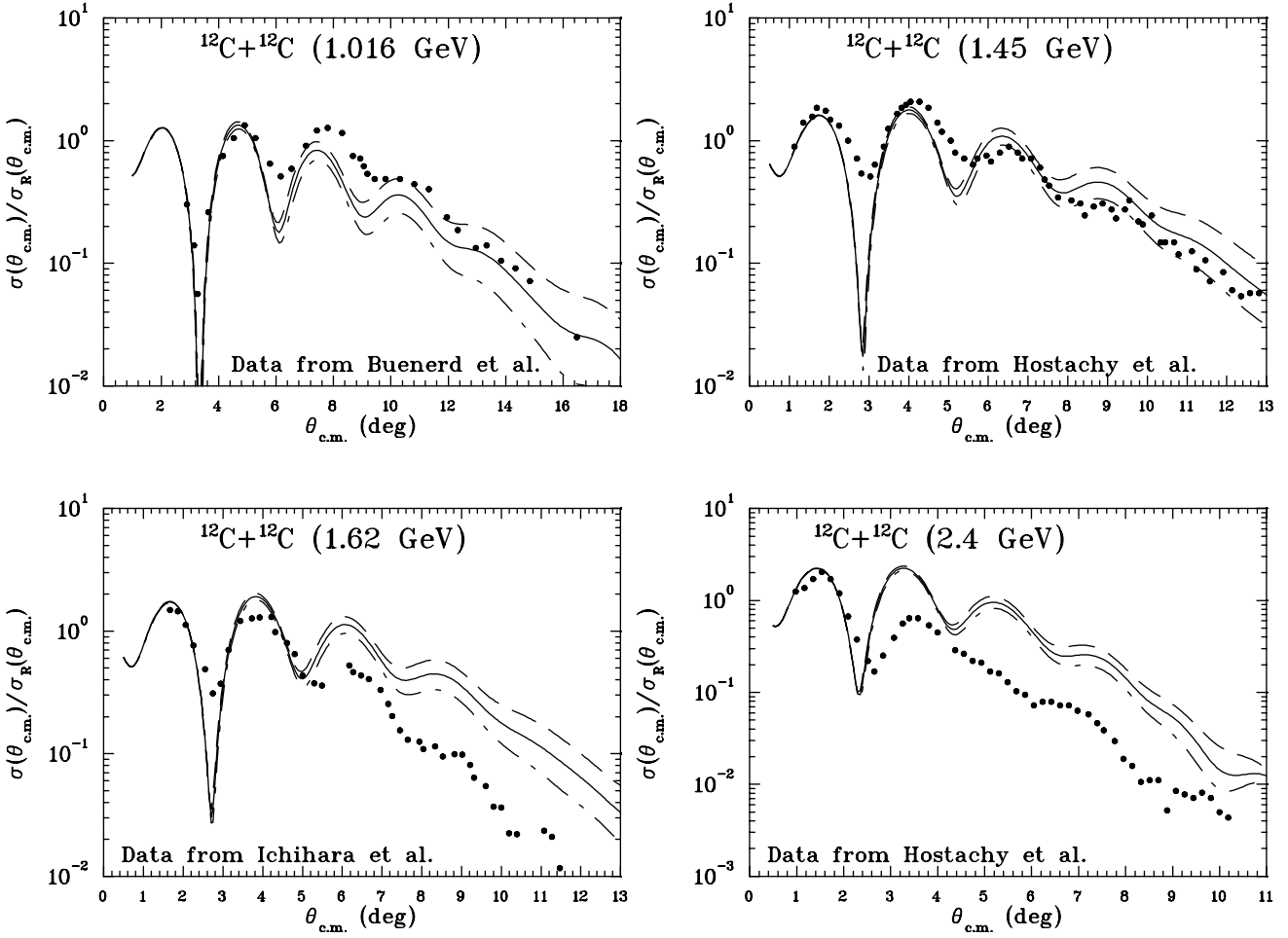,angle=90,height=5in}
\caption{Comparison of three deformations of the carbon nucleus
at four energies. The dashed curves correspond to a spherical 
nucleus (n=0), the solid curve to a ratio of 1.14 (N=2) and the
dash-dot curve to a ratio of 1.73 (n=4). The data are from
Buenerd et al. \cite{buenerd}, Hostachy et al. \cite{hostachy} and
Ichihara et al. \cite{ichihara}.}
\label{defc124}
\end{figure*}

We have carried out a test of the size of the center-of-mass effect in much the 
same way as was done for 
the variational wave function in $\alpha-\alpha$ scattering earlier. The 
$^{12}$C-$^{12}$C scattering calculations just presented were made by 
repeating the realization (in a Monte Carlo loop) of two carbon nuclei and 
then carrying out the evaluation of the necessary equations. The calculation 
can also be done by first constructing a pool of nuclei, each properly 
centered about the center of mass (one million were used in the current 
calculation) and then drawing complete nuclei randomly from this pool for 
each carbon nucleus. These two methods give the same result.

One can now modify the calculation, in the same manner as before, so as to 
choose the 12 different vector coordinates for each carbon nucleus from 12 
different realizations in the pool. In this way one is guaranteed to have 
the same single particle density but with uncorrelated particles.  The 
result of a calculation for this energy with a spherical density for 
$^{12}$C is shown in Fig. \ref{cc1.016cm} where it is seen that both the 
Coulomb and center-of-mass corrections play a large role. The agreement is 
good up to an angle of about 5 degrees ($t\approx -0.045$\ (GeV/c)$^2$) but 
poorer after that.

The result without center-of-mass corrections is shown as the dashed curve. 
While this curve is shown as the ratio to Rutherford as a function 
of angle and Fig. \ref{m4420m1mnc} for helium gives the absolute 
cross section as a function of t, one can see that the effect is 
very similar in the two cases. Thus, at least for this very limited 
sample of two cases, the center-of-mass effect does not decrease.

Figure \ref{bpool} shows the profile function for $^{12}$C for the 
cases with and without CMC. One can understand the behavior to some 
extent. For small values of $b$ the many factors in the product in 
Eq. \ref{basicpf}, (144 for the case of $^{12}$C-$^{12}$C 
scattering), a large fraction 
of which have magnitude less that unity, will cause it to be very 
small. This very small correction to unity leaves the profile 
function at essentially one in this region of $b$, hence is 
insensitive to the CMC.

On the other hand, for large $b$ one can expand the product in 
terms of single, double, triple etc. scatterings. Since a double 
scattering will have two factors of the Gaussian function, the 
triple scattering will have three factors etc., the single 
scattering will dominate for large $b$. Since we are holding the 
single particle density in the c.m. fixed, the CMC will have no 
effect in this region. Because the large $b$ values dominate the 
forward scattering we must expect the small angle cross section to 
be insensitive to the CMC as is observed.

Thus, it is only in a relatively small region of values of $b$ that 
the effect will be influential. From Fig. \ref{bpool} this is for 
$3.5 < b <5.5$ fm. Since, as one decreases $b$ from the external 
region the sensitivity to the CMC will become greater as the number 
of scatterings goes up until this number gets to be so large that 
the product becomes very small in magnitude. 

It is generally believed that carbon has a strong oblate 
deformation. Lesniak and Lesniak \cite{lesniaks} included this 
effect in proton-carbon scattering using Glauber theory. While they 
developed the proper, fully quantum, theory, in the end they used a 
semi-classical approximation simply averaging the amplitude over 
rotated densities (see also Ref. \cite{degliatti}). 
Some recent studies (of fission and reactions \cite{ccfusion}) 
have treated the problem in the same way. As pointed 
out in Ref. \cite{lesniaks}, the electron scattering (at least in 
the single interaction approximation) should not depend on the 
deformation so the average density remains the same as before. We 
assume a form for the density symmetric about the z-axis and with a 
distribution in the polar angle given by

\eq
\rho(r,\theta)\propto \rho_0(r)\sin^{n}\theta.\label{sinn}
\qe

This simple form is inspired by considerations from Ref. \cite{brown} 
(page 62). The paper of Svenne and Mackintosh \cite{svenne} 
presented arguments why $^{12}$C was known to be 
deformed in response to the paper by Friar and Negele \cite{friar} 
who pointed out that, with the usual form of the deformed density, 
the existence of a deformation for $^{12}$C was contrary to the 
electron scattering measurements since the fall off of the density 
in the surface region was strongly affected. The form of Eq. 
\ref{sinn} does not suffer from this problem as can be seen by 
taking the example of $n=2$. In this case we have

$$
\rho(r,\theta)\propto \rho_0(r)\sin^2\theta.\label{sin2}
\propto \frac{2}{3}\rho_0(r)[1-P_2(\cos\theta)]
$$
\eq
=
\rho_0(r)\left(\frac{2}{3}-\frac{8\pi}{15}
\sum_m Y_2^m(\theta_p,\phi_p)Y_2^{m*}(\theta_a,\phi_a)\right)
\qe
where the Legendre polynomial has been expanded in terms of angles 
relative to a fixed axis. 
Here  ($\theta_p,\phi_p$) are angles of a given nucleon in the nucleus
with respect to a fixed axis and the angles ($\theta_a,\phi_a$)
are those of the body symmetry axis relative to the same 
fixed axis. With a one-body operator the second term will give a 
$j_2(qr)$ transform of the density but when the average over the 
direction of the body axis of the carbon nucleus is taken it will 
vanish. Hence a one-body operator probes only the ``spherical 
part'' of the density.

For $n=2$ the probability in $\theta$ is given by
\eq
\rho(\theta)=\frac{3}{4}\sin^2\theta
\qe
which leads to
$$
<z^2>=\frac{1}{5}<r^2>
$$
$$
<x^2>=<y^2>=\frac{2}{5}<r^2>
$$
\eq
<x^2>/<z^2>=2
\qe
For $n=4$ the probability in $\theta$ is given by
\eq
\rho(\theta)=\frac{15}{16}\sin^4\theta
\qe

$$
<z^2>=\frac{1}{7}<r^2>;\ \ <x^2>=<y^2>=\frac{3}{7}<r^2>
$$
\eq
<x^2>/<z^2>=3
\qe
   
These densities look more like donuts than ellipsoids of revolution 
(assuming that this density is applied to all the nucleons and not 
just the p shell). We note that $\alpha$ cluster models would also lead 
to densities with zero at the center. We can estimate the $\beta_2$ 
parameter describing deformation by using prescriptions given by 
Hagino \cite{hagino} based on the rms radii.
   
\begin{figure}[htb]
\epsfig{file=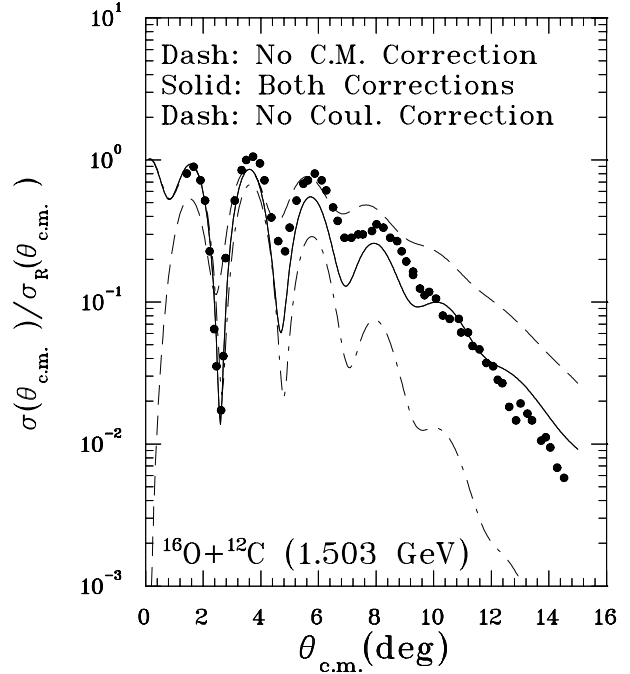,height=3.5in}
\caption{The $^{16}$O-$^{12}$C scattering cross section showing the effect of
the Coulomb and center-of-mass corrections. The 
data are from Roussel-Chomaz et al. \cite{oxygendat}} 
\label{oc1}
\end{figure}

We take
\eq
\gamma\equiv \frac{<x^2>^{\h}}{<z^2>^{\h}}=\frac{1-\h \beta_2
\sqrt{\frac{5}{4\pi}}}{1+\beta_2 \sqrt{\frac{5}{4\pi}}}
\qe
so that
\eq
\beta_2=\frac{1-\gamma}{\h+\gamma}\sqrt{\frac{4\pi}{5}}
\qe

For $n=2$ we have $\beta_2=-0.34$ and for $n=4$ $\beta_2= -0.52$.  
Svenne and Mackintosh \cite{svenne} gave a survey of values obtained 
for $\beta_2$. From their list we see that the Nilson model gives 
$\beta_2=-0.64$ and the $\alpha$ cluster model $\beta=-0.41$. A 
little later Vermeer et al. \cite{vermeer} measured the quadrupole 
moment of the 2$^+$ state from which one can infer a quadrupole 
moment for the ground state of $-22 \pm 10\ e\ $fm$^2$ implying a 
value of $\beta_2=-0.57$ with a 50\% error. In a recent calculation 
of $^{12}$C-$^{12}$C fusion  \cite{ccfusion} the value 
$\beta_2=-0.4$ was used (along with a hexadecapole term) so these 
forms lead to reasonable values of the deformation.

The results for n=2 and n=4 are shown in Fig. \ref{defc124} and 
compared with data. Again it is seen that agreement with data is 
good in the forward direction but deteriorates rapidly at larger 
angles. The deformation plays a significant role at larger angles.

\begin{figure}[htb]
\epsfig{file=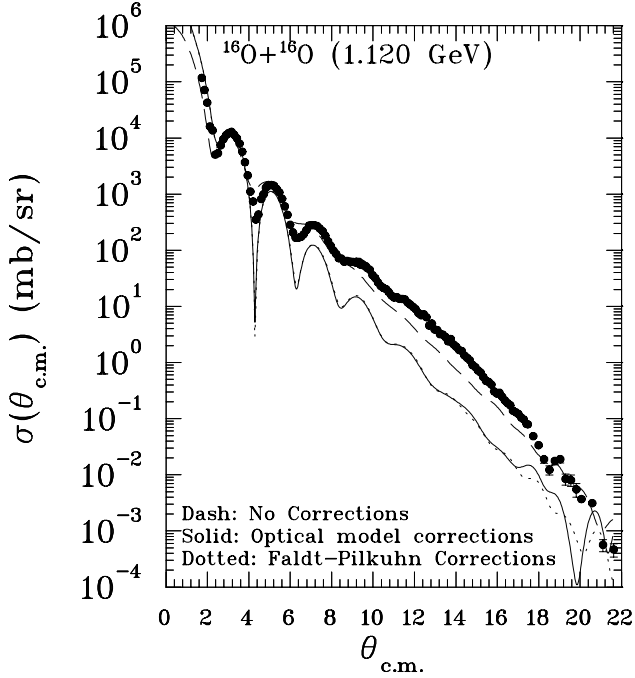,height=3.5in}
\caption{$^{16}$O-$^{16}$O scattering at 1.12 GeV. The data are 
from Nuoffer et al. \cite{nuoffer}, and Khoa et al. \cite{khoa}\label{ooabs}}
\end{figure}

\begin{figure}[htb]
\epsfig{file=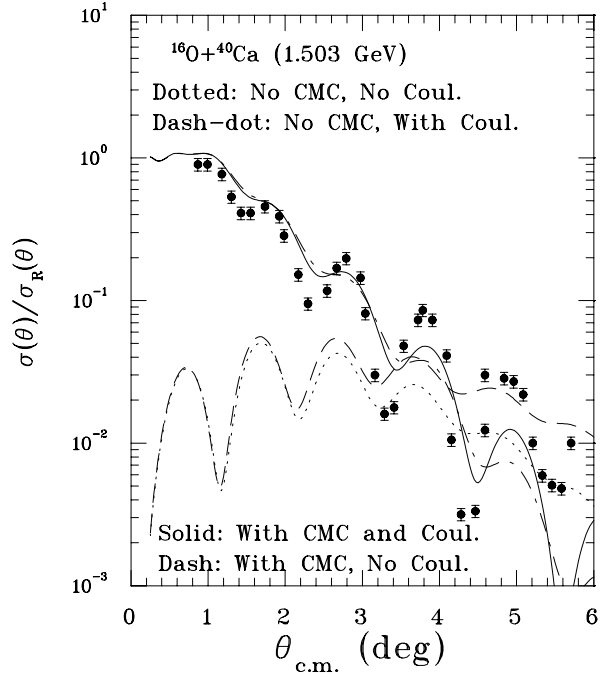,height=3.5in}
\caption{The $^{16}$O-$^{40}$Ca scattering cross section showing the effect of 
the Coulomb and c.m. corrections. The data are from Roussel-Chomaz et al. \cite{oxygendat}}
\label{oxca}
\end{figure}

\subsection{$^{16}$O-$^{12}$C Scattering\label{ocsection}}

The scattering of Oxygen on Carbon is shown in Fig. \ref{oc1}. The 
effect of the correction for the Coulomb interaction is quite large.  For the 
same scattering the effect of the correction 
for the center of mass is also shown. It is seen 
that the agreement a very forward angles is very good with the 
Coulomb correction playing a substantial role. While it cannot be 
said that the agreement with data is good at larger angles, the 
calculation follows the trend of the data quite well which is not 
true if either the Coulomb or center-of-mass correction is not 
included.

\subsection{$^{16}$O-$^{16}$O  Scattering\label{oosection}}
Using the density found earlier we now calculate
Oxygen scattering from Oxygen at 1.120 GeV. The comparison with
the data of Nuoffer et al. \cite{nuoffer} is shown in Fig. 
\ref{ooabs}. For angles beyond 5 degrees there is a considerable
discrepancy (more after the Coulomb correction) but inside of that 
angle the agreement is quite good. 

In order to test the inner Coulomb correction we made an optical-model
fit to the data \cite{nuoffer}. A simple form was used
\eq
V_{\rm 
opt}(r)=-\frac{V_0}{1+e^{(r-r_R)/a_R}}-\frac{iW_0}{1+e^{(r-r_I)/a_I}}
\qe
with $r_R=2r_R^0(16)^{\frac{1}{3}}$ and 
$r_I=2r_I^0(16)^{\frac{1}{3}}$. A uniform charge density with a 
radius $r_q=2.\times 1.3\times (16)^{\frac{1}{3}}$ was used. The 
fit parameters were $V_0=150.4$ MeV, $r^0_R=0.784$ fm, $a_R=0.897$ 
fm, $W_0=44.4$ MeV, $r^0_I=1.005$ fm, and $a_I=0.745$ fm. By 
calculating with and without the Coulomb interaction, a Coulomb 
correction can be obtained by the method shown in Eq. \ref{cdef}. 
Alternatively, we can make the correction with the method F\"aldt 
and Pilkuhn \cite{faldt} by modifying the integral over the profile 
function. As seen in Fig. \ref{inner} the results are nearly 
identical over most of the angular range.

\subsection{$^{16}$O-$^{40}$Ca Scattering\label{oxcasection}}

The result of the $^{16}$O-$^{40}$Ca scattering is shown in Fig. 
\ref{oxca} and compared with data of Roussel-Chomaz et 
al. \cite{oxygendat}. Again one sees the importance of the 
center-of-mass and Coulomb corrections. The agreement at forward 
angles is moderately good but is considerably worse at larger 
angles. For another treatment of this reaction at this energy
see Ref. \cite{chakim}.

\begin{figure}[htb]
\epsfig{file=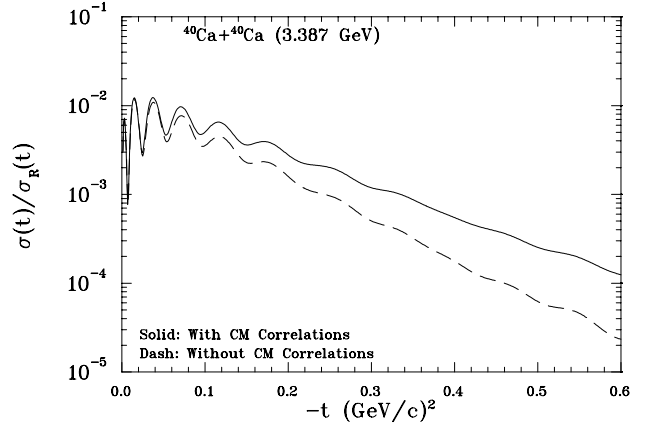,angle=90,height=2.2in}
\caption{The ratio of the $^{40}$Ca-$^{40}$Ca cross section to
the Rutherford cross section at 3.387 GeV using
the appropriate parameters from Ref. \cite{arndt} for this energy. 
\label{casig3.387}}
\end{figure}

\subsection{$^{40}$Ca-$^{40}$Ca scattering\label{cacasection}}

Figure \ref{casig3.387} shows the cross section for 
$^{40}$Ca-$^{40}$Ca scattering at a kinetic energy of 3.387 GeV. 
There is no data to compare with and if there were they would be 
dominated by pure Coulomb. In this case there are 1600 factors in 
the product in Eq. \ref{gtilde} which, if expanded in multiple 
scattering as was done in the paper by Franco and Yin \cite{fy} 
would give $2^{1600}\approx4\times 10^{481}$ terms. The calculation 
is shown to illustrate the fact that, even for these medium-heavy nuclei, 
the center-of-mass correction remains important. We see that the 
effect of the CMC is not very different from the carbon-carbon 
scattering and is as large as a factor of 5. Since the basic CMC 
effect is the order of 1/40 in this case it is important to 
understand how many times it enters in to the calculation. For $b$ 
small the profile function is unity and only a relatively few 
factors are needed to achieve this result.

\subsection{$^6$He-$^{12}$C scattering\label{he6csection}}

\begin{figure}[htb] 
\epsfig{file=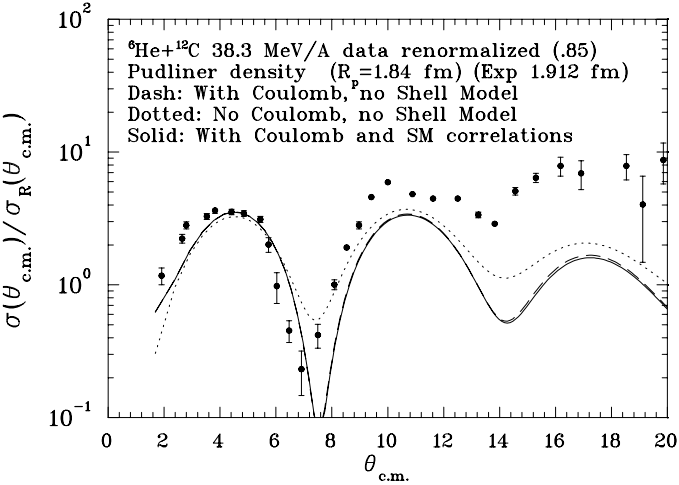,angle=90,height=2.5in} 
\caption{$^6$He-$^{12}$C scattering at 38.3 MeV/A using the $^6$He 
density from Pudliner et al. \cite{pudliner}. The data are from 
Lapoux et al. \cite{lapoux}. The experimental determination of the
proton radius is from Ref. \cite{wang}.
} \label{he6corr38.3} \end{figure}

\begin{figure}[htb] 
\epsfig{file=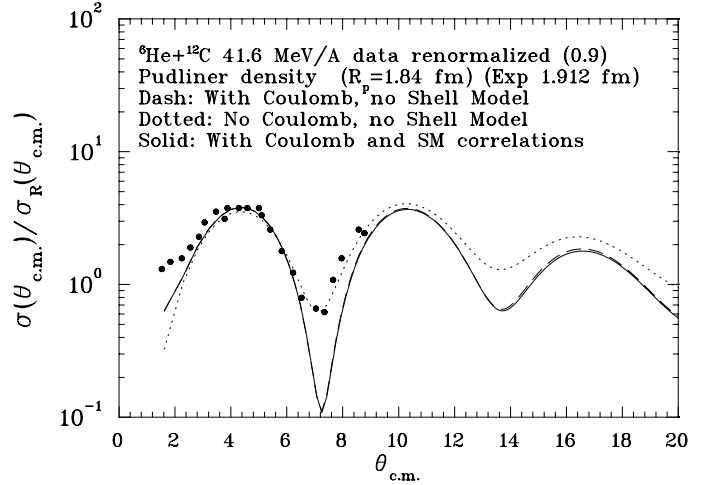,angle=90,height=2.5in} 
\caption{$^6$He-$^{12}$C scattering at 41.6 MeV/A using the $^6$He 
density from Pudliner et al. \cite{pudliner}. The data are from 
Al-Khalili et al. \cite{he6c1241.6}} \label{he6corr41.6} 
\end{figure}

The Pudliner density for $^6$He \cite{pudliner} was represented
by two auxiliary functions (one for protons and one for neutrons 
following Eqs. \ref{proteq} and \ref{neuteq}.
The proton auxiliary density was given by
\eq
r^2\rho_p(r)=0.273776 \frac{r^{15}b_p^{16}e^{-b_pr}}{15!}
+0.726224\frac{r^8c_p^9e^{-c_pr}}{8!}
\qe
with
\eq b_p=14.3466\ {\rm fm}^{-1}\ {\rm and}\ c_p=4.7125\ {\rm fm}^{-1}
\qe
and the neutron auxiliary density by
\eq
r^2\rho_n(r)=0.208969\frac{r^{14}e_ne^{15}e^{-e_nr}}{14!}
+0.791031\frac{r^2f_n^3e^{-f_nr}}{2!}
\qe
where
\eq
e_n=5.72753\ {\rm fm}^{-1}\ {\rm and}\ f_n=1.65566\ {\rm fm}^{-1}.
\qe
A ``shell model'' correlation was (optionally) included by coupling two 
$j=\frac{3}{2}$ neutrons to zero total angular momentum which 
results in a factor in the density of

\eq
\h (1+\cos^2\theta_R)
\qe
where $\theta_R$ is the relative angle between the two valence 
neutrons. This factor was carried as a weight in the Monte Carlo 
calculation so the calculations with and without the factor can be 
done in the same run. Figures \ref{he6corr38.3} and 
\ref{he6corr41.6} show the results for the calculation of 
$^6$He-$^{12}$C scattering. It is seen that the effect of 
the shell-model correlation is very small. The Coulomb correction plays a 
significant role even for these low-Z nuclei. Perhaps a part of the 
reason for this is that the correlation is only among two of the 
six particles.

\section{Discussion\label{discussion}}

The  Glauber approximation can fail for a number of reasons not
necessarily associated with the basic assumptions. These 
reasons include:

\begin{itemize}
\item At low energies the Fermi motion may cause significant
corrections to the fixed-nucleon approximation.

\item The double spin flip in the nucleon-nucleon interaction
which has been neglected may require a significant correction
if the isospin constraints are not sufficient to eliminate it.

\item The single spin flip (occurring an even number of times) may become 
important for sufficiently large angles.

\item At high energies the coherent production of mesons
constitutes an additional inelastic channel which is beyond
the Glauber approximation. 

\item At small impact parameters there should be significant corrections to the 
nucleon-nucleon interactions because of the higher nuclear densities.

\item Correlations from the shell model are of the same range as those
from the center of mass and may play a role as the nuclear penetration
becomes greater.
 
\end{itemize} 

\begin{figure}[htb]
\epsfig{file=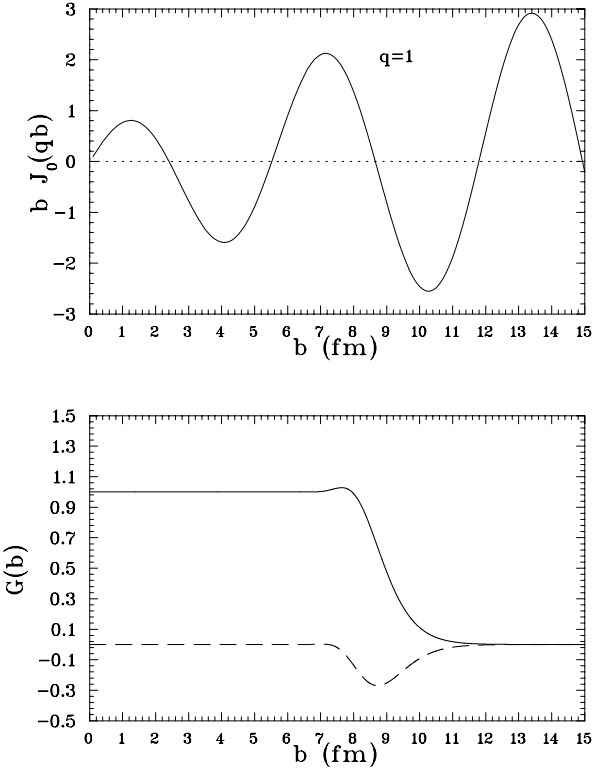,height=4in}
\caption{The factor of the kernel $b J_0(qb)$ which represents the 
weighting of the nuclear profile function (top panel) and the 
nuclear profile function for $\alpha-^{208}$Pb (lower panel).}
\label{j0pic} \end{figure}

In looking over the comparisons with data presented in the
paper it is seen that often there is a forward region where
the prediction is moderately good (sometimes quite good) followed
by a more-or-less sudden transition to a poor agreement. Perhaps
the best example of this is the scattering of $\alpha$-particles
from $^{208}$Pb (see Fig. \ref{alpb}). The transition is seen
to take place at 8$^o$ at 288 MeV, 7$^o$ at 340 MeV, 5$^o$ at
480 MeV and 4.5$^o$ at 699 MeV. These angles and the corresponding
momentum transfer and values of -t are shown in Table \ref{table2}.
All of the transition points are seen to correspond to a momentum 
transfer of about 1 fm$^{-1}$ [--t=0.04 (GeV/c)$^2$]. For the 
$^{16}$O-$^{12}$C scattering (see Fig. \ref{oc1}) the transition
angle is about 4$^o$ which corresponds to a momentum transfer of
1.01 fm$^{-1}$. For $^{16}$O-$^{16}$O scattering (Fig. \ref{ooabs})
the transition angle is around 5$^o$, corresponding to a momentum
transfer of 1.28 fm$^{-1}$. For the $^{16}$O-$^{40}$Ca scattering
(Fig. \ref{oxca}) the transition angle is about 2.5$^o$ which
corresponds to a momentum transfer of 1.06 fm$^{-1}$. The remainder
of the cases studied are shown in Table \ref{table2} as well and show
a similar effect.

Figure \ref{j0pic} shows that at $q_C=1$ fm$^{-1}$ for the momentum transfer
the interior of the nucleus is sampled. So $q_C=1$ fm$^{-1}$ does not seem to
correspond to the onset of the nuclear matter being probed.
\begin{table*}[ht]
\begin{center}
\begin{tabular}{|clccccc|}
\hline
\hline
Case& T$_{Lab}$ (GeV) [Ref.]&$\theta_C$( 
deg)&$q_C$(fm$^{-1})$&$q_C^2$ 
(GeV/c)$^2$&$\sqrt{s}-m_T-m_P $(GeV)&Figure\\
\hline
$\alpha$-$^{208}$Pb& 0.288 
 \cite{bonin}&8&1.04&0.042&0.28&\ref{alpb}\\ 
$\alpha$-$^{208}$Pb &0.340 
\cite{bonin}&7&0.99&0.038&0.33&\ref{alpb}\\ 
$\alpha$-$^{208}$Pb 
&0.480 \cite{bonin}&5&0.85&0.028&0.47&\ref{alpb}\\ 
$\alpha$-$^{208}$Pb& 0.699 
\cite{bonin}&4.5&0.93&0.034&0.65&\ref{alpb}\\ 
$^{16}$O-$^{12}$C &1.503 
\cite{oxygendat}&4&1.02&0.041&0.64&\ref{oc1}\\
$^{16}$O-$^{16}$O &1.120 
\cite{nuoffer,khoa}&5&1.28&0.064&0.55&\ref{ooabs}\\
$^{16}$O-$^{40}$Ca &1.503 
\cite{oxygendat}&2.5&1.06&0.044&1.06&\ref{oxca}\\ 
$^{12}$C-$^{12}$C &1.016 \cite{buenerd}&4.5&0.96&0.036&0.50&\ref{cc1.016cm}\\
$\alpha$-$^{16}$O &0.240 \cite{liu}&7&0.66&0.017&0.19&\ref{alox240}\\
$\alpha$-$^{16}$O &0.400 \cite{wakasa}&7&0.89&0.031&0.32&\ref{alox400}\\ 
$^{12}$C-$^{12}$C &1.449 \cite{hostachy}&4&1.01&0.040&0.71&\ref{defc124}\\
$^{12}$C-$^{12}$C &1.620 \cite{ichihara}&3&0.80&0.025&0.80&\ref{defc124}\\
$^{12}$C-$^{12}$C &2.400 \cite{hostachy}&2&0.66&0.017&1.17&\ref{defc124}\\
$\alpha$-$^{40}$Ca& 1.37 \cite{alphacadat}&8&2.18&0.184&1.23&\ref{alphaca}\\ 
$\alpha$-$^{12}$C& 4.20 \cite{morschzp,morschprc}&&&&2.87&\ref{alphac7gevcrho}\\ 
$\alpha$-$^{12}$C &1.37 \cite{chaumeaux}&&&&0.99&\ref{vpar1.37ai}\\ 
$\alpha$-$\alpha$& 2.55 \cite{berger}&&&&1.18&\ref{m4420m1mnc}\\
$\alpha$-$\alpha$& 4.20 \cite{satta}&&&&1.87&\ref{alphaalpha7gev/c}\\ 
\hline
\hline
\end{tabular}
\end{center}
\caption{Critical values of the angles and their equivalents in 
momentum transfer. Also given is the energy in the center of mass 
available for coherent meson production. The last four cases have no 
critical values listed because all data points available have a 
momentum transfer greater $\ge$ 1 fm$^{-1}$.} The $^6$He-$^{12}$C
cases have not been included since it is not possible to claim agreement
in the forward direction without absolute data.
\label{table2}\end{table*}

\vspace{.5cm}

Acknowledgments

We thank Dr. H. G. Bohlen for providing tables of the $^{16}$O-$^{16}$O 
data and Dr. V. Lapoux for communicating to us the $^6$He-$^{12}$C 
data at 38.2 MeV. WRG thanks the Laboratoire de Physique Nucl\'{e}aire et de 
Hautes \'{E}nergies for support during visits when part of this work was done.

\begin{appendix}

\section{Partial-wave projection}

We make the partial wave projection from the formula
\eq
\frac{1}{2ik}\sum_{\ell=0}^{\infty}(2\ell+1)(S_{\ell}-1)P_{\ell}
(\cos\theta)
=ik\int_0^{\infty}bdbJ_0(qb)G(b).
\qe
where $q^2=2k^2(1-\cos\theta)$. In order to make this projection it is
useful to have the expansion
\eq
J_0(qb)=\sum_{\ell=0}^{\infty}h_{\ell}(kb)P_{\ell}(\cos\theta)
\label{j0expan}.\qe
From Ref \cite{as} 9.1.79 we have
\eq
J_0(qb)=J_0^2(kb)+2\sum_{n=1}^{\infty}J_n^2(kb)\cos n\theta.
\qe
In order to get the desired expression we need the expansion
\eq
\cos 
n\theta=\sum_{\ell=0}^{\infty}a_{n,\ell}P_{\ell}(\cos\theta)
\label{cosn},\qe
which actually cuts off at $\ell=n$ as we shall see. 
With the use of DeMoivre's theorem on can see
\eq
\cos n\theta=\h\sum_{m=0}^n 
\left(\begin{array}{c}m\\n\end{array}\right)
[i^m+(-i)^m]\sin^m\theta \cos^{n-m}\theta
\qe
$$
=\sum_{m=even}^n \left(\begin{array}{c}m\\n\end{array}\right)
(-1)^{\frac{m}{2}}\sin^m\theta \cos^{n-m}\theta
$$
where $\left(\begin{array}{c}m\\n\end{array}\right)$ is the
binomial coefficient.
With $k=m/2$
$$
=\sum_{k=0}^{\frac{n}{2}} 
\left(\begin{array}{c}2k\\n\end{array}\right)
(-1)^k(\sin^2\theta)^k \cos^{n-2k}\theta
$$
$$=\sum_{k=0}^{\frac{n}{2}} 
\left(\begin{array}{c}2k\\n\end{array}\right)
(-1)^k(1-\cos^2\theta)^k \cos^{n-2k}\theta
$$ 
$$
=\sum_{k=0}^{\frac{n}{2}}\sum_{j=0}^k
 \left(\begin{array}{c}2k\\n\end{array}\right)
(-1)^k(-1)^j\left(\begin{array}{c}j\\k\end{array}\right)
 \cos^{n-2k+2j}\theta.
$$
With $m=n-2k+2j$ so that $m=n,n-2,n-4,\dots$
$$=\sum
 \left(\begin{array}{c}n-m-2j\\n\end{array}\right)
(-1)^{\frac{n-m}{2}}\left(\begin{array}{c}j\\\frac{n-m}{2}-j\end{array}
\right)
\cos^m\theta
$$
$$
=\sum c_{n,m}\cos^m\theta.
$$
Thus we see that $\cos n\theta$ can be expanded in powers of
$\cos\theta$ with powers of the same parity as $n$ with maximum
power $n$. We can now use the integral of a Legendre polynomial
over a power of $x$ (Ref. \cite{as} 8.14.15) 
\eq
\int_{-1}^1P_{\ell}(x)x^m dx=\frac{\pi^{\h}2^{-m}\Gamma(1+m)}
{\Gamma(1+\h m-\h \ell)\Gamma(\h\ell+\h 
m+\frac{3}{2})}=d_{m,\ell}.
\qe
Note that this integral is zero if $\ell$ and $m$ are not of the 
same parity or if $\ell>m$.

We can write the expression for the $a_{n,\ell}$ as
\eq
a_{n,\ell}=(\ell+\h)\sum_{\ell\le m\le n,m-n\ 
even}c_{n,m}d_{m,\ell}.
\qe

This expression was coded and the coefficients calculated 
in this way; it works but the terms for different powers of
$\cos\theta$ get very large and cancel in the sum so, even with 
double precision, one is limited to $n$ of the order of 35 for 
small values of $\ell$ which is not enough for our problem. The values
for $n=\ell$ and other values of $\ell$ close to $n$ are always
calculated correctly. A useful check is
\eq
\sum_{\ell=0}^na_{n,\ell}=1
\qe
which follows from Eq. \ref{cosn} with $\theta=0$.
   
However, there is a much easier way to get the coefficients 
needed. Using the trigonometric identity
\eq
\cos n\theta=2\cos(n-1)\theta\cos\theta-\cos(n-2)\theta
\qe
and the recursion relation for the Legendre polynomials
\eq
xP_{\ell}(x)=\frac{\ell P_{\ell-1}(x)+(\ell+1)P_{\ell+1}(x)}
{2\ell+1}
\qe
we find
$$
\int_0^{\pi} \cos n\theta P_{\ell}(\cos\theta)\sin\theta d\theta
\equiv b_{n,\ell}
$$
\eq
=\frac{2}{2\ell+1}\left[
\ell b_{n-1,\ell-1}+(\ell+1)b_{n-1,\ell+1}\right]-b_{n-2,\ell}.
\qe
Since $a_{n,\ell}=\frac{2\ell+1}{2}b_{n,\ell}$ we have
\eq
a_{n,\ell}=\frac{2\ell a_{n-1,\ell-1}}{2\ell-1}
+\frac{2(\ell+1)a_{n-1,\ell+1}}{2\ell+3}-a_{n-2,\ell}\label{recur}
\qe
with the conditions
\eq
a_{0,0}=1;\ \ \ a_{1,1}=1;
\qe
and
\eq
a_{n,\ell}=0
\qe
if $n$ and $\ell$ are not of the same parity or if $\ell>n$. All
values of these coefficients can be quickly calculated using the
recursion relation \ref{recur} without 
any numerical problem. Using these values then
\eq
h_0(kb)=J_0^2(kb)+2\sum_{n=1}^{\infty}J_n^2(kb)a_{n,0}  
\qe
\eq
h_{\ell}(kb)=2\sum_{n=\ell}^{\infty}J_n^2(kb)a_{n,\ell};\ \ 
\ell>0
\qe
$J_0(x)$ and $J_1(x)$ are calculated to one part in $10^7$ and the 
recursion relation for Bessel functions is used for fixed $b$ to obtain
the higher values of $n$. Just beyond $n=bk$, the 
magnitude of $J_n(x)$ decreases rapidly and the recursion relation 
fails shortly thereafter. The recursion is cut off when $J_n^2(x)$ is 
less than $10^{-6}$.

The functions $h_{\ell}(kb)$ have some interesting and potentially
useful properties. For example starting from Eq. \ref{j0expan} with
$$
\int_0^{\infty}J_0(qb)J_0(q'b)bdb=\delta^{(2)}(q-q')
$$
\eq=\frac{\delta(q-q')}{q}
=2 \delta(q^2-q'^2)=\frac{1}{k^2}\delta(x-x')
\qe
one can show that
\eq
\int_0^{\infty}h_{\ell}(y)h_{\ell'}(y)ydy=\frac{2\ell+1}{2}\delta_{\ell,\ell'}
\qe

Once the functions $h_{\ell}(bk)$ have been found, the process can
be reversed, i.e. we can find the nuclear profile function for any
given set of matrix elements. First from the inverse Hankel transform
we have:
\eq
G(b)=\frac{1}{ik}\int_0^{\infty} F(q)J_0(qb) q dq
\qe
Since the amplitude is assumed to fall off rapidly we can replace the
upper limit with a finite, but large, one, namely $2k$ so that
\eq
G(b)\approx \frac{1}{ik}\int_0^{2k} F(q)J_0(qb) q dq
\qe
Changing the integration variable to $u\equiv q^2$ we
have 
\eq
G(b)\approx \frac{1}{2ik}\int_0^{4k^2} F(\sqrt{u})J_0(qb) du
\qe
Now changing the integration variable to $x=\cos\theta$
we have
\eq
G(b)\approx \frac{2k^2}{2ik}\int_{-1}^1 A(x)J_0(q(x)b) dx  
\qe
where $A(x)=F[q(x)]$ and
\eq
A(x)=\frac{1}{2ik}\sum_{\ell=0}^{\infty}(2\ell+1)(S_{\ell}-1)P_{\ell}(x)
\qe
Using Eq. \ref{j0expan} we have
\eq
G(b)\approx -\sum_{\ell=0}^{\infty}(S_{\ell}-1)h_{\ell}(kb)  
\qe

\section{Variational density for $^4$He\label{varga}}

One can solve for the $^4$He wave function by Monte Carlo Green's 
Function methods \cite{carlson,pudliner}. In this case the walkers either 
represent a probability proportional to the wave function itself 
(hence the walkers do not give the density needed for our 
scattering problem) or, if one uses importance sampling then the 
walkers represent the product of the trial wave function and the 
true wave function and thus is more suitable for representing the 
density. In the work here we will use a variational trial wave 
function where the walkers represent directly the density to 
calculate $\alpha-\alpha$ scattering.

Varga et al. \cite{varga} performed Monte Carlo calculations using a Metropolis 
sampling of a $^4$He wave function obtained with Green's Function Monte Carlo 
Methods. We will make a comparison between the method given previously and a 
simplified version of this type of calculation based on the variational 
algorithm.

The variational method operates with the estimator of the energy
given by
$$
E_T=
\frac {\int d\bfR \psi^*_T(\bfR)(T+V)\psi_T(\bfR)}{\int d\bfR 
\psi^*_T(\bfR) \psi_T(\bfR)}
$$
\vspace*{-.2in}
\eq
=\frac {\int d\bfR \rho(\bfR)(\frac{T\psi_T(\bfR)}{\psi_T(\bfR)}+V)
}{\int d\bfR \rho(\bfR)}
\qe
where $\bfR$ represents the set of coordinates which describe the 
system. Using the Metropolis algorithm to represent the density,
it is automatically normalized. The trial wave function is
assumed to depend on some number of parameters and, upon varying these
parameters the minimum energy achieved is guaranteed to be greater than
or equal to the true ground state energy of the system. The trial
wave function normally is expected to give a good representation of
the true wave function but there is no guarantee of that.

In order to calculate with a realistic wave function we use a trial
wave function which results from the following variational calculation for $^4$He.
Since the true wave function must be translationally invariant it can
depend only on the six relative coordinates, $\bfr_{ij}=\bfr_i-\bfr_j$.

We choose the form
\eq
\psi(\bfr_1,\bfr_2,\bfr_3,\bfr_4)=\prod_{j>i}f(r_{ij})
\qe
where $r_{ij}=|\bfr_{ij}|$ and the function, $f(r)$ is arbitrary at this
point. It will be chosen with a number of parameters to be selected to
minimize the energy. We can evaluate the kinetic energy needed as
\begin{widetext}
\eq
\sum_{i=1}^4\frac{\nabla_i^2\psi(\bfr_1,\bfr_2,\bfr_3,\bfr_4)}{
\psi(\bfr_1,\bfr_2,\bfr_3,\bfr_4)}=2\sum_{j>i}\left(\frac{f''(r_{ij})}
{f(r_{ij})}+2\frac{f'(r_{ij})}{r_{ij}f(r_{ij})}\right)+2\sum_{[i,j,k]}
\bfr_{ij}\cdot\bfr_{ik}\frac{f'(r_{ij})f'(r_{ik})}
{r_{ij}r_{ik}f(r_{ij})f(r_{ik})}
\qe
\end{widetext}
 The values of the three indices in the last sum are: [123], [124], 
[134], [213], [214], [234], [312], [314], [324], [412], [413] and [423].
They can be obtained with nested loops 
\eq
\{i=1,4 [j=1,3; j\ne i (k=j+1,4; k\ne i)]\}
\qe
With such a form it is relatively easy to calculate the kinetic 
energy since one only needs to be able to calculate the function, 
$f$, and its first and second derivatives. The derivatives can be 
calculated numerically if need be.

We took for the form of $f$
\eq
f(r)=(1-e^{-cr})\frac{e^{-ar}}{b+r}.
\qe
The first factor vanishes at the origin and provides the effect of 
a (very mild) repulsive correlation. The rest of the function has 
the proper asymptotic form with the constant $b$ avoiding the 
singularity at the origin.

We used a modified version \cite{book} the Malfliet-Tjon \cite{mt} 
potentials since we are neglecting spin in this calculation. The 
variational calculation over-binds $^4$He by about 7 MeV (35 MeV 
instead of 28 MeV) so we increased the strength of the repulsive 
part of the modified Malfliet-Tjon potentials (by 6\%) in order to 
have a more realistic binding.

The variational calculation for the energy was carried out using 100,000 walkers. 
It was found that the energy was not very sensitive to the 
parameter $b$ so it was fixed at 1.0 fm.  Then, for various fixed 
values of $c$, the energy was calculated as a function of $a$. For 
the fixed values of $b$ and $c$ the rms radius is a unique 
function of $a$. We used $a=0.203 $\ fm$^{-1}$ and $c=1.0$\ 
fm$^{-1}$. The rms radius is a crucial parameter in the scattering 
and we obtained a value of 1.44 fm. The calculation was repeated
10 times to give a collection of one million walkers to use in the
scattering calculations involving the $\alpha$ particle.

The coordinates needed for the scattering calculation were taken from the 
Metropolis algorithm every 2000 steps to be certain that each realization was 
independent. One million realizations of the helium density were calculated and 
written out to files. When the scattering was calculated, the coordinates of each 
nucleus were drawn from a nuclear representation chosen randomly from the pool of 
one million possible nuclear realizations.

\end{appendix}

 \end{document}